\documentclass[sigconf]{acmart}
\AtBeginDocument{%
  }

\usepackage{graphicx}
\usepackage[inline]{enumitem} %
\usepackage{multicol}
\usepackage{amsmath}
\usepackage{threeparttable}
\usepackage{dblfloatfix}
\usepackage{float}
\usepackage{tabularx}
\usepackage{booktabs}
\usepackage{ragged2e,enumitem,multirow}
\usepackage{graphicx}
\usepackage{siunitx}
\usepackage{threeparttable}
\usepackage{pifont}
\usepackage{comment}
\usepackage{makecell,adjustbox}
\usepackage{pgfplots}
\usepackage{xurl} 
\usepackage{caption}
\usepackage{tcolorbox}
\usepackage{listings}
\usepackage{ragged2e} 
\usepackage{arydshln}
\usepackage[dvipsnames]{xcolor}

\newcommand{\summary}[1]{\smallskip\noindent\textbf{#1.}}
\newcolumntype{P}[1]{>{\RaggedRight\arraybackslash}p{#1}} 

\newcommand{\pquote}[1]{\textit{``#1''}}

\usepackage{array}

\newcommand{\one}{\ding{182}} 
\newcommand{\two}{\ding{183}} 
\newcommand{\three}{\ding{184}} 
\newlist{compactitem}{itemize}{5}
\setlist[compactitem]{leftmargin=*, nosep}
\setlist[compactitem, 1]{label=\textbullet}
\setlist[compactitem, 2]{label=\textendash}
\setlist[compactitem, 3]{label=\textasteriskcentered}
\setlist[compactitem, 4]{label=\textperiodcentered}

\makeatletter
\let\orgItem\item
\NewDocumentCommand\fixedItem{ o }{%
   \IfNoValueTF{#1}%
      {\orgItem}%
      {\orgItem[#1]\def\@currentlabel{#1}}%
}
\makeatother

\newlist{questions}{enumerate}{3}
\setlist[questions]{align=left, labelwidth=2em, labelsep=.5em, listparindent=0pt, itemindent=0pt, leftmargin=!, before=\let\item\fixedItem}
\setlist[questions, 1]{labelindent=0pt, label=\textbf{Q\arabic*}, widest=99}
\setlist[questions, 2]{labelindent=-2.5em, label*=\textbf{\_\Alph*}, widest=26}
\setlist[questions, 3]{labelindent=-2.5em, label*=\textbf{\_\roman*}, widest=9}

\newlist{answers}{itemize}{1}
\setlist[answers]{leftmargin=*, nosep, align=left, label=$\bigcirc$}
\newlist{answers*}{itemize*}{1}
\setlist[answers*]{label=$\bigcirc$}

\newcommand{\fillinline}{{[\textit{\small{open text response}}]}}

\newenvironment{packed_itemize}{
    \begin{list}{\labelitemi}{\leftmargin=1em}
    \setlength{\itemsep}{1pt}
    \setlength{\parskip}{0pt}
    \setlength{\parsep}{0pt}
    \setlength{\headsep}{0pt}
    \setlength{\topskip}{0pt}
    \setlength{\topmargin}{0pt}
    \setlength{\topsep}{0pt}
    \setlength{\partopsep}{0pt}
}{\end{list}}

\copyrightyear{2026}
\acmYear{2026}
\setcopyright{cc}
\setcctype{by}
\acmConference[CCS '26]{Proceedings of the 2026 ACM SIGSAC
Conference on Computer and Communications Security}
{November 15--19, 2026}{The Hague, Netherlands.}
\acmBooktitle{Proceedings of the 2026 ACM SIGSAC Conference
on Computer and Communications Security (CCS '26), November
15--19, 2026, The Hague, Netherlands}
\acmISBN{979-8-4007-2871-6/2026/11}
\acmDOI{10.1145/3830454.3846694}
\begin{document}

\date{}

\title[Do Hackers Dream of Electric Teachers?]{Do Hackers Dream of Electric Teachers?: A Large-Scale, In-Situ Measurement of Cybersecurity Student Behaviors and Educational Performance with AI Tutors}
\titlenote{Extended version of the CCS~'26 paper (doi:10.1145/3830454.3846694), adding appendices omitted from the conference version for space.}

\author{Michael Tompkins}
\affiliation{%
  \institution{Arizona State University}
  \city{Tempe}
  \state{Arizona}
  \country{USA}
}
\email{mctompk1@asu.edu}

\author{Nihaarika Agarwal}
\affiliation{%
  \institution{Arizona State University}
  \city{Tempe}
  \state{Arizona}
  \country{USA}
}
\email{nagarw48@asu.edu}

\author{Ananta Soneji}
\affiliation{%
  \institution{Arizona State University}
  \city{Tempe}
  \state{Arizona}
  \country{USA}
}
\email{asoneji@asu.edu}

\author{Robert Wasinger}
\affiliation{%
  \institution{Arizona State University}
  \city{Tempe}
  \state{Arizona}
  \country{USA}
}
\email{rwasinge@asu.edu}

\author{Connor Nelson}
\affiliation{%
  \institution{Arizona State University}
  \city{Tempe}
  \state{Arizona}
  \country{USA}
}
\email{connor.d.nelson@asu.edu}

\author{Kevin Leach}
\affiliation{%
  \institution{Vanderbilt University}
  \city{Nashville}
  \state{Tennessee}
  \country{USA}
}
\email{kevin.leach@vanderbilt.edu}

\author{Rakibul Hasan}
\affiliation{%
  \institution{Arizona State University}
  \city{Tempe}
  \state{Arizona}
  \country{USA}
}
\email{rakibul.hasan@asu.edu}

\author{Adam Doup\'e}
\affiliation{%
  \institution{Arizona State University}
  \city{Tempe}
  \state{Arizona}
  \country{USA}
}
\email{doupe@asu.edu}

\author{Daniel Votipka}
\affiliation{%
  \institution{Tufts University}
  \city{Medford}
  \state{Massachusetts}
  \country{USA}
}
\email{dvotipka@cs.tufts.edu}

\author{Yan Shoshitaishvili}
\affiliation{%
  \institution{Arizona State University}
  \city{Tempe}
  \state{Arizona}
  \country{USA}
}
\email{yans@asu.edu}

\author{Jaron Mink}
\affiliation{%
  \institution{Arizona State University}
  \city{Tempe}
  \state{Arizona}
  \country{USA}
}
\email{jaron.mink@asu.edu}

\renewcommand{\shortauthors}{Tompkins et al.}

\ccsdesc[500]{Applied computing~Computer-assisted instruction}
\ccsdesc[300]{Social and professional topics~Computing education}
\ccsdesc[100]{Security and privacy~Social aspects of security and privacy}

\keywords{AI tutors; LLMs; cybersecurity education; help-seeking behavior; conversation analysis; computing education}

\begin{abstract}
To meet the ever-increasing demands of the cybersecurity workforce, AI tutors have been proposed for personalized, scalable education. But, while AI tutors have shown promise in introductory programming courses, no work has evaluated their use in hands-on exploration and exploitation exercises (e.g., ``Capture the Flag'') commonly used to teach cybersecurity. 
In particular, it is unclear how students use AI tutors, or what types of use correlate with greater success in solving the challenges in real, large-scale cybersecurity courses.

To answer this, we conducted a semester-long observational study of an embedded AI tutor with 309 students in an upper-division introductory cybersecurity course.
By analyzing 142,526 student queries sent to the AI tutor across 383 cybersecurity challenges spanning 9 core cybersecurity topics and an accompanying end-of-semester survey, we find (1) what queries and conversation styles students use with AI tutors, (2) how these styles relate to challenge completion, and (3) students' perceptions of AI tutors in cybersecurity education.
In particular, we identify three broad AI tutor conversation styles among students: Short (bounded, few-turn exchanges),  Reactive (repeatedly submitting code and errors), and Proactive (driving problem-solving through targeted inquiry). 
We also find that these styles are significantly correlated with challenge completion, and that the completion-rate gap between styles widens as materials become more advanced.
Furthermore, students valued the tutor's availability but reported that it became less useful for harder material.
Based on our results, we provide suggestions for security educators and developers on practical AI tutor use.
\end{abstract}

\maketitle

\section{Introduction}

Cybersecurity experts are essential to maintaining the security of organizations, yet the cybersecurity workforce demand is consistently unmet~\cite{isc2workforce2024}.
Bridging this gap is hard in part because cybersecurity training is fundamentally hands-on, with the current trend consisting of practice-based learning such as Capture the Flag (CTF) challenges~\cite{votipka2018hackers} and structured vulnerability discovery exercises~\cite{votipka2021hacked, Nelson2024PwnLearning}.

Because of their effectiveness, university curricula are increasingly adopting practice-based approaches in which students solve progressively difficult  challenges~\cite{neslon2024DOJO,nelson2026FiveYears,Nelson2024PwnLearning, vykopal2020CTF}; however, two obstacles make practice-based approaches difficult to scale.
\textit{First}, these exercises are often open-ended explorations of complex systems; this  
can be particularly demanding for learners as they must navigate unfamiliar tools, complex system internals, and open-ended problems with minimal structure~\cite{votipka2021hacked, MatteiJGI2025}, which can lead to discouragement or attrition~\cite{MatteiJGI2025,FultonV4A2023}. %
\textit{Second}, prior work has shown that mentoring help from experts is important for overcoming these challenges~\cite{votipka2018hackers,FultonV4A2023}; %
however, formal courses often lack the teaching assistant (TA) capacity to provide individual mentoring at scale, and students may hesitate to approach authority figures~\cite{FultonV4A2023}.

In response to such concerns, AI tutoring has emerged as a proposed solution to provide individual student mentoring at scale.
In contrast to general chatbots, which educators increasingly worry are used to bypass students' productive struggle rather than support learning~\cite{bastani2025GenAI, jost2024impact, marwan2020unproductive},
purpose-built AI tutors have been proposed as an alternative, and introductory CS deployments report that personalized feedback and self-paced progress benefit students~\cite{xiao2024quicktahelp,kazemitabaar2024codeaid, yang2024debugging, sheese2024helpseeking}.
However, it remains unclear whether these benefits transfer to cybersecurity education given its exploratory nature, looser structure, and cross-domain reasoning.
More specifically, we ask:
\begin{enumerate}[leftmargin=*, label=\textbf{RQ\arabic*}]
\item How do students interact with AI tutors while solving cybersecurity challenges? \label{rq:interaction}
\item \label{rq:performance}How do AI tutor conversations relate to challenge performance, and how does this vary by cybersecurity topic? 
\item What do students believe are the benefits and drawbacks of AI tutors in cybersecurity courses?  \label{rq:student-beliefs}
\end{enumerate}
To answer these questions, 
we conducted a semester-long observational
study of an embedded AI tutor with 309 students in an upper-division introductory cybersecurity course. By analyzing 142,526 student queries sent to the AI tutor across 383 cybersecurity challenges spanning 9 core cybersecurity topics and an accompanying
end-of-semester survey on their opinions of the system, we provide insights on the effectiveness of cybersecurity AI tutors in practice.

\textit{First}, we find that participants commonly ask seven types of queries, and perhaps surprisingly, only a small minority of queries (3.5\%) are direct solution requests.
\textit{Second}, using this taxonomy of queries to fit a mixture hidden Markov model, we find that students' conversations naturally partition into three broad styles: Short (bounded, single-purpose exchanges), Reactive (tending to begin with a confused student), and Proactive (tending to begin with focused, specific questions).
Compared to Proactive conversations, Reactive conversations are twice as likely to contain a direct solution request (25.8\% vs.\ 11.8\%) and six times as likely to contain more than one (10.9\% vs.\ 1.8\%).
\textit{Third}, we find that conversation style is associated with challenge completion: Proactive conversations show equal or higher completion rates than Reactive ones in all nine modules, and the gap between styles widens from under $1\%$ on introductory material to $14.3\%$ on the most advanced.
\textit{Fourth}, from students' survey responses we find that they valued the tutor's constant availability, but reported that it became less useful as material grew harder.

Based on these findings, we provide a set of recommendations for practical and effective AI tutor usage for educators, identify opportunities for improvements in AI tutor system designs, and discuss how these results can be interpreted within educational theory.

\section{Background \& Related Work}
\label{sec:related-work}

\summary{LLM-Assisted Coding} \label{llm_assisted_coding}
Deployments of large language model (LLM) assistants into introductory programming and systems courses have seen students use them for code generation, debugging, and explanation, while highlighting the importance of tool design to preserve student agency~\cite{kazemitabaar2024codeaid, yang2024debugging, sheese2024helpseeking}. Jo{\v{s}}t et al.~\cite{jost2024impact} link real-world LLM usage patterns to outcomes, finding that reliance on code generation and debugging may correlate with lower performance, while explanation-focused use appears less harmful. 
Bastani et al.~\cite{bastani2025GenAI} provide causal evidence for this concern through a large-scale study where students given unrestricted access to GPT-4 performed worse on unassisted exams than students with no access.

\summary{AI Tutors in CS Education}
Many systems now treat LLMs as conversational tutors rather than simple code generators. Studies of these systems emphasize opportunity for personalized feedback and guidance while warning about over-reliance and academic integrity~\cite{cambaz2024use}.
Several deployments integrate prompt construction into curriculum, framing prompt design and refinement as learning objectives~\cite{denny2023conversing, promptprogramming2024dialogue}, while others design and deploy LLMs as conversational tutors for CS coursework, enabling multi-turn conversational assistance instead of one-shot queries~\cite{perez2024dt4coding, xiao2024quicktahelp}. A growing body of research also suggests that the design of these systems matters more than the underlying model, with evidence that AI tutor design can mitigate learning harms from unrestricted LLM access~\cite{bastani2025GenAI}. A common element is the use of Socratic-style questioning rather than direct answers, encouraging students to reason through problems rather than passively receive solutions~\cite{favero2024enhancing, ding2024boosting}. However, no consensus exists for which designs are best. %

\summary{Help-Seeking in Computing Education}
The way students seek help has long been recognized as a predictor of learning outcomes, with Karabenick and Gonida~\cite{karabenick2018academic} highlighting adaptive learners who build targeted understanding, executive learners who seek help to complete without engaging, and avoidant learners who avoid help despite needing it.
In computing education, these manifest in observable ways: students group into distinct help-seeking profiles by which resources they use~\cite{battestilli2022academic}, students whose forum questions show their reasoning earn higher course grades~\cite{vellukunnel2017complementing}, and students in hint-based environments exhibit both help abuse (requesting hints without thinking) and help avoidance (struggling without requesting hints), both of which can impede learning~\cite{marwan2020unproductive}. Recent work extends these findings to LLM-based tools, showing that students who prompt for conceptual understanding perform better on assignments and exams~\cite{mcnichols2025studychat}. However, no study has examined this in cybersecurity education, where tasks are more open-ended and span multiple knowledge domains, or whether the style--outcome associations observed in other contexts reflect effects of interaction strategy or selection of which students adopt each style.

\begin{figure*}[t]
\centering
\includegraphics[width=\linewidth]{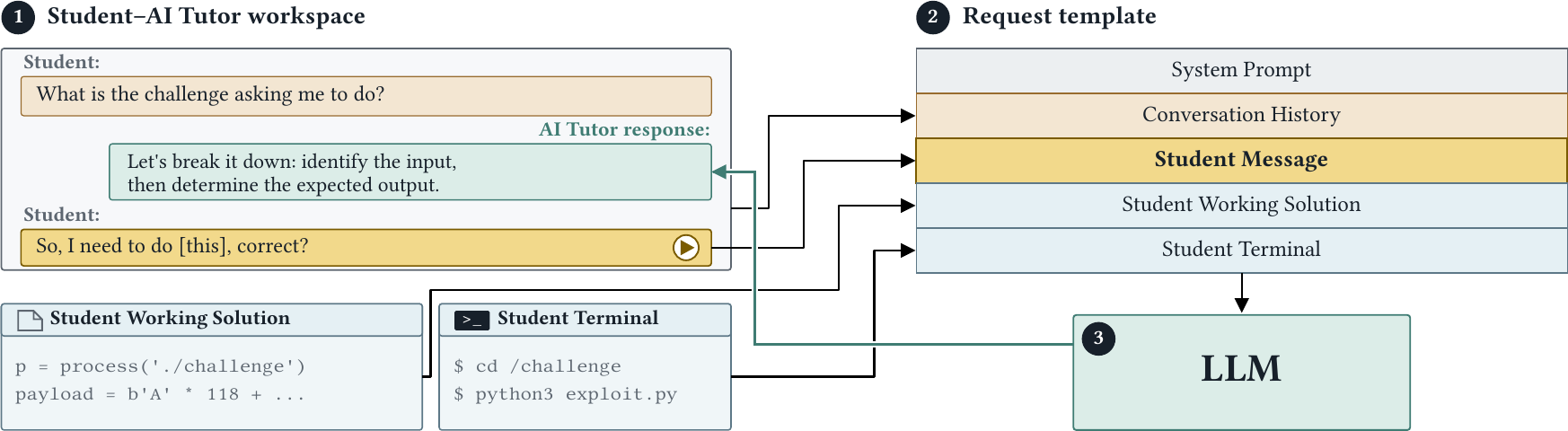}
\Description{Diagram of the AI tutor architecture. A student's chat message, the conversation history, and captured terminal and solution-file context are bundled into a templated request and sent to the LLM, whose response returns to the chat interface.}
\caption{
\textbf{AI Tutor Overview} -- \textmd{\small Participants solve challenges inside an instrumented container, capturing their active terminal and file context. When a student interacts with the AI tutor (\one{}), the system bundles up the AI tutor's system prompt, the participant's conversation history and query, and the captured context from the container (\two{}). This gets sent to the LLM (\three{}) which processes the query and responds to the participant.
}}
\label{fig:llm-tutor-arch}
\end{figure*}

\summary{Cybersecurity Education}
Unlike general programming education, cybersecurity education centers on hands-on application across diverse knowledge domains.
Yet, existing work on generative AI in these classrooms largely focuses on high-level pedagogy and self-directed learning~\cite{marquardson2024embracing, elkhodr2025generative, Triplett2025AI}, with fewer efforts building dedicated tutoring systems~\cite{mukherjee2025generative, nelson2025sensai} or giving students direct LLM access for security tasks~\cite{wang2024teaching}.
Thus, while prior work has begun to integrate generative AI into cybersecurity education and reports on students' perceptions of AI tutors, no work focuses on how students use and perform with these systems.

\section{Methodology}
\label{sec:methodology}

To understand how students use AI tutors in cybersecurity education, we integrated SENSAI~\cite{nelson2025sensai}, an AI tutoring system, into a cybersecurity course at a public US university in the spring of 2025 and assessed students’ AI tutor conversations, course performance, and perceptions of the AI tutor.
Our IRB-approved processes ensured compliance with FERPA regulations,
fairness in grading, and general standards of ethical research practices (see Appendix~\ref{appendix:ethics-extended}).

\begin{table}[t!]
\small
\setlength{\tabcolsep}{3pt}
\renewcommand{\arraystretch}{1.0}
\centering
\begin{tabular}{ll}
\toprule
\textbf{Module} & \textbf{Topics} \\
\midrule
1. Linux & Linux command line interface \\
2. Data \& Access & Encodings (ASCII, Base64) and permissions \\
3. Web \& SQL & HTTP and SQL queries, inspecting traffic \\
4. Web Security & Path traversals, XSS, command injections \\
5. Comp. Arch. & x86-64 assembly, memory, debugging \\
6. Net. Security & Traffic analysis, firewalls, DoS, TCP/UDP \\
7. Cryptography & Symmetric/asymmetric, hashing, DHKE \\
8. Reverse Engr. & Iterative reversing of a file format parser \\
9. Binary Security & Memory corruption, buffer overflows \\
\bottomrule
\end{tabular}
\caption{\textbf{Cybersecurity Course Modules}}%
\label{tab:module_defs}
\end{table}

\summary{Course Content and Challenges} We study a 15-week online cybersecurity course in which students access all course materials through pwn.college~\cite{neslon2024DOJO, nelson2026FiveYears}.
The course is divided into nine modules (see Table~\ref{tab:module_defs}) covering commonly taught security concepts:%
\footnote{We removed a tenth capstone module due to low participation caused by course grading policies, i.e., students could earn high grades without completion.}%
the course begins by teaching the Linux command line and file access control, and then moves to cover a range of security topics such as web security, cryptography, and binary exploitation.
Each module consists of pre-recorded lecture videos and 21 to 92 (avg. 44) CTF-style challenges students solve for course credit. In total, students are assigned 396 challenges.

Each challenge contains two components: (1)~a description of tasks students must solve to complete the challenge that decreases in detail and direction as the challenges increase in difficulty,
and (2)~a Dockerized Linux system that hides a unique text string (a ``flag'').\footnote{All flags come in a predefined format (e.g., ``flag\{XXXX.XX\}'').} 
To complete the challenge, students apply concepts taught in the lectures in the Linux environment to access the flag and submit it to the platform. 
For example, a web security challenge 
requires students to perform a command injection attack against a vulnerable HTTP endpoint to retrieve the flag (\S\ref{appendix:example_chall_desc}).
These challenges made up 80\% of the course grade.

\summary{AI Tutor Architecture and Workflow} \phantomsection\label{tutor_arch}
Our AI tutor was designed to assist students when solving challenges by answering conceptual questions and providing light coding assistance.
As shown in Figure~\ref{fig:llm-tutor-arch}, the architecture of the tutor consists of 
\one{} a chat-based interface for student queries,
\two{} a request template that combines 
(a) a system prompt instructing the LLM to be a Socratic mentor,
(b) the conversation history between the tutor and a student,
(c) the new student query, and
(d) the current Linux system context, composed of the latest saved file and active terminal history to capture command usage and outputs,
and \three{} an LLM to which the templated request is sent and from which a response is received; we specifically use GPT-4o's API as it was the state-of-the-art model at the time. %
We chose to instruct our LLM to be a Socratic mentor based on emerging evidence that guided questioning produces better learning outcomes than direct answers~\cite{favero2024enhancing,bastani2025GenAI} and best practices for educational prompt templates~\cite{ding2024boosting, chang2023prompting}. The full system prompt is provided in \S\ref{appendix:sys_prompt}. The included Linux system context simulates a TA's ability to ``look over the shoulder,'' enabling the model to provide targeted guidance without requiring the student to describe their working state.
In addition, the system \textit{does not} have access to a solution, discouraging extraction and encouraging students to build one cooperatively.

At any time during a challenge, students can open the chat interface and submit free-form queries to the AI tutor.
By default, all elements shown in \two{} in Figure~\ref{fig:llm-tutor-arch} are included when a query is made, but can be optionally removed via a toggle button (which, in practice, was rarely used; \S\ref{appendix:context-toggles}).
When a query is submitted, the request is bundled, sent to the LLM, and the LLM's response is presented to the student. Students may submit additional queries as needed.
The tutor's context window is not preserved across challenges or challenge restarts.

\subsection{Observation Data}
\label{sec:observation_data}
We collected data on students' (1) tutor conversations, (2) challenge completion, and (3) 
perceptions of the AI tutor.

\summary{AI Tutor Queries}
Participants interacted with the system by sending queries. A \textit{query} consists of the participant's message submitted to the tutor with its accompanying system context.
A \textit{response} is the message the AI tutor provides back to the student. 
The complete sequence of queries and responses between a student and the AI tutor for a given challenge makes up a \textit{conversation}.
Once a student moves on to the next challenge, a new conversation begins.
For our analysis, if a student restarts a challenge or returns to it after working on others, we merge these sessions into a single conversation (i.e., students can have at most one conversation per challenge). %
We collected participating students' conversations throughout the semester, resulting in at least one conversation for $383$ of the $396$ assigned challenges.

\summary{Challenge Completions}
Participating students' \emph{performance} (completion rate) for all challenges is captured throughout the semester.

\summary{End-of-Semester Survey} 
We conducted an end-of-semester survey to understand and measure students' perception of the AI tutor (\S\ref{appendix:eos-survey}).
The survey contained
five sections with both open- and closed-ended questions about their background knowledge (Q1--Q3),
perceptions of utility (Q4--Q9),
system usability (Q10--Q13),
comparisons to TAs (Q14--Q17),
and possible system improvements (Q18).

\subsection{Qualitative Data Analysis}
\label{sec:qualitative_analysis}
To understand student-tutor conversations and how students perceive our AI tutor, we qualitatively coded students' queries to the AI tutor
and open-ended responses to the end-of-semester survey. %

\summary{AI Tutor Queries}
To analyze student queries, we used an iterative open coding approach~\cite{strauss1998basics} with coders familiar with the course and the cybersecurity concepts to code students' queries.\footnote{Because some queries are only classifiable in context of the full conversation, the coder had access to the tutor-visible system context and the tutor's response.}
The first author read a random sample of 500 queries (from the full 142,526 query set) to construct the initial codebook. 
This initial codebook was then explained to a second coder (another author),
and two coders independently coded queries in rounds of 500.
After each round, the coders met to calculate inter-rater reliability (IRR) for each code. 
We calculated IRR using Cohen's $\kappa$, as it is a conservative measure that accounts for chance agreement~\cite{mchugh2012kappa}.
Coders resolved disagreements and adjusted the codebook to refine code definitions or add codes as necessary.
After 8 rounds of independent coding (4{,}000 queries total), they reached a high level of agreement across codes ($\kappa=.84$)~\cite{mchugh2012kappa}, producing the final codebook.

Due to the scale of our data, we chose to use an LLM (GPT-5.2~\cite{openai2025gpt52card}) to apply our finalized codebook to the full set of 142,526 queries.
To code each query, we submitted a prompt containing strict code definitions and few-shot examples of correct coding to the LLM, along with the target query, tutor-accessible system context, and the tutor's response.
Following best practices for LLM-based coding~\cite{Hou2024PromptDeductiveCoding, Liu2025QualitativeCoding}, we iteratively refined this prompt by comparing it against a ground-truth set of 1{,}500 queries from the previously manually coded set. We randomly sampled 500 queries and had the LLM code them, then calculated the IRR between the LLM's codes and our ground-truth coding using Cohen's $\kappa$, treating the LLM as an additional coder measured against human consensus. Two coders reviewed disagreements and iteratively revised the prompt. This process was repeated for 8 rounds 
until the LLM demonstrated high overall agreement ($\kappa=.82$)~\cite{mchugh2012kappa}.\footnote{We randomly sample from our 1{,}500-query ground truth each round. While this can introduce bias for human coders, we simply delete the LLM's context between rounds.} 
The full codebook, including the $\kappa$ values for each task family, is provided in Table~\ref{tab:query_codebook}. 
We then used the refined LLM prompt to code the entire dataset that served as input into our quantitative analyses, specifically the mixture hidden Markov model (MHMM) conversation-style analysis (\S\ref{sec:mhmm_clusters}) and mixed-effects logistic regression (\S\ref{sec:statistical_analysis}).

\summary{End-of-Semester Survey}
To analyze the five open-response reflection questions (Q8, Q9, Q13, Q17, Q18), we used an inductive open coding approach~\cite{strauss1998basics}.
Two coders independently coded student reflections in rounds of 150 responses, randomly selected from the 293 total survey responses. The coders met between rounds to resolve disagreements and update the codebook as necessary. 
After 5 rounds, the coders achieved Cohen's $\kappa$ above $.80$ for all codes. The primary coder then coded the %
remaining responses. 
The final codebook is provided in Table~\ref{tab:reflection_codebook_definitions}.

\subsection{Quantitative Data Analysis}
Using the coded student queries, we first evaluate how patterns in conversations emerge.
To do this, we fit MHMMs~\cite{helske2019seqhmm} to the coded query sequences (see \S\ref{sec:mhmm_clusters} for details).
After finding these conversational patterns, or ``conversation styles'', we then evaluated how they relate to performance by modeling challenge completion
as a function of conversation style and the module being attempted. %
To do this, we fit mixed-effects logistic regression
using whether the participant completed the challenge associated with each conversation as our dependent variable. We then fit participants' conversation styles and module as fixed effects, and each participant as a random effect (see \S\ref{sec:statistical_analysis} for details).

\subsection{Eligibility, Recruitment, and Consent}
\label{subsection:eligibility-recruitment}
All students enrolled in the course were eligible for study participation. To enroll, students must have completed prerequisite courses in basic computer architecture, programming languages, and data structures and algorithms, and must be on-campus students.
Prior to using the tutor, students had to consent to have their conversation logs collected and analyzed. 
An additional, separate consent form was also provided for the end-of-semester survey, which enabled the collection and association of survey responses, LLM conversations, and students' course performance.
Only students who (1) opted into the AI tutor's data collection, 
(2) completed the end-of-semester survey and opted to share their responses, and (3) had at least one conversation with the AI tutor in the studied modules participated in our study.
If students chose to opt out of using the AI tutor, they were still able to access and use other course resources.

\begin{table}[t]
\small
\setlength{\tabcolsep}{3pt}
\renewcommand{\arraystretch}{1.0}
\centering
\begin{tabular}{@{}l r@{}}
\toprule
& Count (\%) \\
\midrule
\multicolumn{2}{@{}l}{\textbf{Prior Cybersecurity Understanding}} \\
\hspace{3mm} Not at All Knowledgeable    & 135 (46.1) \\
\hspace{3mm} Slightly Knowledgeable      &  92 (31.4) \\
\hspace{3mm} Somewhat Knowledgeable      &  31 (10.6) \\
\hspace{3mm} Moderately Knowledgeable    &  25 (8.5) \\
\hspace{3mm} Extremely Knowledgeable     &  10 (3.4) \\
\midrule
\multicolumn{2}{@{}l}{\textbf{Frequency of AI Use (outside course)}} \\
\hspace{3mm} Never                              &   8 (2.7) \\
\hspace{3mm} Rarely ($\leq$1 a week)            &  61 (20.8) \\
\hspace{3mm} Sometimes (2--3 times a week)      & 126 (43.0) \\
\hspace{3mm} Often (Once a day)                 &  55 (18.8) \\
\hspace{3mm} Very Often (More than once a day)  &  43 (14.7) \\
\bottomrule
\end{tabular}
\caption{\textbf{Participants' Cybersecurity Knowledge and AI Use} -- 
\textmd{\small Self-reported background knowledge and AI use of the 293 participants (of 309) included in the survey analysis (Q1--Q2).}
}
\label{table:participant-background}
\end{table}

\subsection{Participants}
In total, $309$ of the $720$ enrolled students participated. Each participant completed the course, used the AI tutor at least once, and submitted an end-of-semester survey.
Within the end-of-semester survey, $16$ of our $309$ participants self-reported never having used the tutor (despite their logged conversations); we therefore only exclude them from our survey analysis.
As shown in Table~\ref{table:participant-background}, of these $293$ participants, most had little prior cybersecurity education (Q1), with 77\% ($n$=$227$) describing themselves as ``Not at all knowledgeable'' or ``Slightly knowledgeable'' of cybersecurity concepts. This level of cybersecurity knowledge is expected of introductory cybersecurity course students, our target demographic.
Most participants were active AI users, with 76\% ($n$=$224$) reporting they use AI at least 2--3 times a week for problem-solving, and only 3\% ($n$=$8$) reported never using AI,
which aligns with the levels of college student AI use found in prior work~\cite{Hou2025GenAI}.

\subsection{Limitations}
\label{sec:limitations}
In providing and analyzing real-world student conversations and educational performance, our study also has several limitations.
\textit{First}, while an observational study of real-world AI tutor-student interactions provides high ecological validity, we cannot separate correlation from causation, and thus cannot conclude whether conversation styles cause better outcomes, or whether students with better outcomes approach conversations differently.
This stems from necessary ethical constraints: studying AI tutor use by real students at scale means we cannot experimentally assign or deny access.
While we run additional analysis decomposing this association into within- and between-student components (\S\ref{appendix:mundlak}), future work should ethically confirm these findings within a controlled setting.
\textit{Second}, while it is possible that students who opt in to our study differ from the full student population, by comparing participants' per-module challenge completion against aggregate course statistics, we find that our participants' completion closely tracks the full course: median completion differs by at most $4.8$ percentage points in all nine modules (with no consistent bias).
\textit{Third}, while using other AI tools and consulting with other students was against both the study's and the university's academic integrity policies, we are unable to verify whether students used other resources that may have affected their performance.
However, unauthorized resource use is a common issue among educators, and thus, these results are representative of practical course deployments.
\textit{Fourth}, while we use challenge completions as our proxy for performance, they do not necessarily reflect student learning.
Thus, while we cannot strictly claim learning outcomes, these results bear weight on educational outcomes (i.e., grades) commonly accepted across modern educational institutions.
\textit{Fifth}, our use of GPT-4o may introduce model-specific idiosyncrasies; as such, we focus on student behavior (queries to the model, educational performance) rather than how the tutor responds, which we leave to future work.

\section{Student Interactions With the AI Tutor (\ref{rq:interaction})}
\label{ref:desc_stats_intro}
Throughout the semester, 309 participants made 142{,}526 queries across 25{,}261 conversations with the AI tutor. 
Table~\ref{tab:module-usage} shows the distribution of conversations, student adoption rates, and average query count per conversation.\footnote{Query count represents the length of the conversation in queries from the student to the AI tutor, not factoring in the tutor's responses.}
Participants sent a median of 353 queries across 75 conversations, with most conversations being brief (53.7\% contain $\leq$3 queries). Participation declined throughout the course, though 41\% of participants used the tutor in all modules. A per-module breakdown of query families is given in \S\ref{appendix:query-families-by-module}.

\begin{table}[t]
\small
\setlength{\tabcolsep}{3pt}
\renewcommand{\arraystretch}{1.0}
\centering
\begin{tabular}{@{}l l rrr@{}}
\toprule
\textbf{Module} & \textbf{Chal. \#} & \textbf{$n$ (\%)} & \textbf{Convs. (\%)} & \textbf{Avg. C : Q} \\
\midrule
1. Linux & 84 & 258 (83) & 3,802 (15) & 14.7 : 3.8 \\
2. Data \& Access & 38 & 279 (90) & 3,029 (12) & 10.9 : 6.3 \\
3. Web \& SQL & 46 & 268 (87) & 3,899 (15) & 14.5 : 4.5 \\
4. Web Sec. & 27 & 261 (84) & 2,815 (11) & 10.8 : 8.0 \\
5. Comp. Arch. & 65 & 263 (85) & 4,124 (16) & 15.7 : 5.1 \\
6. Network Sec. & 29 & 269 (87) & 3,103 (12) & 11.5 : 6.6 \\
7. Cryptography & 34 & 231 (75) & 2,134 (8) & 9.2 : 6.8 \\
8. Reverse Engr. & 39 & 209 (68) & 1,433 (6) & 6.9 : 5.2 \\
9. Binary Sec. & 21 & 181 (59) & 922 (4) & 5.1 : 6.6 \\
\bottomrule
\end{tabular}
\caption{\textbf{Tutor Usage by Module} 
-- 
\textmd{{\small  
Percentages indicate the proportion of all participants ($n$=$309$) who queried the AI tutor during a given module, and the percentage of total conversations ($m$=$25{,}261$) in a module. ``Avg. C : Q''  is the average number of conversations per participant (C) and the average number of queries per conversation (Q). 
``Chal. \#'' is the count of challenges with $\geq1$ tutor-student conversation. 
}}}

\label{tab:module-usage}%
\end{table}

\subsection{Queries}
\label{sec:queries}
\begin{table}[t]
\small
\setlength{\tabcolsep}{3pt}
\renewcommand{\arraystretch}{1.0}
\centering
\begin{tabular}{@{}l l r@{}}
\toprule
\textbf{Query Type} & \textbf{Subtype} & \textbf{Count (\%)} \\
\midrule

\multirow{2}{*}{%
  \shortstack[l]{Verify \& Fix\\[-2pt]\emph{\scriptsize Total: 49,496 (34.7\%)}}
}
  & Debugging             & 33,325 (23.4) \\
  & Confirmation          & 16,171 (11.3) \\
\midrule

\multirow{2}{*}{%
  \shortstack[l]{Provide Info\\[-2pt]\emph{\scriptsize Total: 31,005 (21.8\%)}}
}
  & Paste Context         & 17,334 (12.2) \\
  & Observations          & 13,671 (9.6) \\
\midrule

\multirow{2}{*}{%
  \shortstack[l]{Implement\\[-2pt]\emph{\scriptsize Total: 24,519 (17.2\%)}}
}
  & Procedure Guidance    & 20,800 (14.6) \\
  & Code Generation       &  3,719 (2.6) \\
\midrule

\multirow{4}{*}{%
  \shortstack[l]{Get Unstuck\\[-2pt]\emph{\scriptsize Total: 18,940 (13.3\%)}}
}
  & Direction Request     &  9,421 (6.6) \\
  & Confusion             &  4,458 (3.1) \\
  & Vague Request         &  2,963 (2.1) \\
  & Help Request          &  2,098 (1.5) \\
\midrule

\multirow{2}{*}{%
  \shortstack[l]{Understand\\[-2pt]\emph{\scriptsize Total: 11,166 (7.8\%)}}
}
  & Concept Guidance      &  8,744 (6.1) \\
  & Challenge Guidance    &  2,422 (1.7) \\
\midrule

\multirow{2}{*}{%
  \shortstack[l]{Solution Request \\[-2pt]\emph{\scriptsize Total: 4,984 (3.5\%)}}
}
  & \multirow{2}{*}{Solution Request} & \multirow{2}{*}{4,984 (3.5)} \\
&&\\
\midrule

\multirow{3}{*}{%
  \shortstack[l]{Peripheral\\[-2pt]\emph{\scriptsize Total: 2,416 (1.7\%)}}
}
  & Social Turn           &  1,361 (1.0) \\
  & Non-Sequitur          &    663 (0.5) \\
  & Course/Platform     &    392 (0.3) \\
\bottomrule

\end{tabular}
\caption{\textbf{Submitted Student Queries}
--
\textmd{{\small  We present the
$m$=$142{,}526$  participant queries made, grouped by type and subtype.
}}}
\label{tab:query-codebook}
\end{table}

\label{axial_taxonomy}
As Table~\ref{tab:query-codebook} shows, participants submitted 16 query subtypes, which we organized via axial coding~\cite{strauss1998basics} into 7 primary query type families based on each subtype's perceived goal. 
Below, we describe each query type (and subtypes as ``[\textbf{subtype}]'').

\summary{Verify \& Fix}
In 34.7\% ($m$=$49{,}496$) of queries,
participants asked the AI tutor to double-check their assumptions and fix issues.
In about two-thirds of these queries ($m$=$33{,}325$), participants used the tutor to fix issues they discovered in code [\textbf{Debugging}].
Participants typically pasted error messages and included file context in their query (see \S\ref{tutor_arch}),
and optionally asked the tutor to help interpret or fix the issue. These could include highly technical issues such as P70 
informing the AI tutor, \pquote{the lea produces an error saying absolute address can not be RIP-relative},
or assistance in using other debugging methods.
In the remaining third of these queries ($m$=$16{,}171$), participants used the tutor to determine whether their challenge approach was correct [\textbf{Confirmation}].
This ranged from simple verification questions, e.g., \pquote{is my [jump] table implementation correct}~(P164), to broadly confirming understanding before implementation, e.g., \pquote{could i first shift right 32 then shift left 64 then move it to rax?}~(P250).

\summary{Provide Info}
In $21.8\%$ ($m$=$31{,}005$) of queries, participants communicated their perception of the state of the problem to the AI tutor. 
In about half of these cases ($m$=$17{,}334$), they directly submitted terminal output or code snippets without any accompanying interpretation or explanation [\textbf{Paste Context}].
In the other half of cases ($m$=$13{,}671$), they provided information not captured by the included context [\textbf{Observations}]; e.g., P238 %
submitted a query observing \pquote{the address for win() is at 0x1d15}. 

\begin{table*}[t]
\setlength{\tabcolsep}{3pt}
\renewcommand{\arraystretch}{1.0}
\centering
\small

\begin{tabular}{@{}r @{\quad} l @{\quad} r @{\quad} l @{\quad} r @{\quad} l @{\quad} r@{}}
\toprule
Rank
& \multicolumn{2}{l}{Single-query ($m$=$6{,}036$)}
& \multicolumn{2}{l}{Two-query ($m$=$4{,}459$)}
& \multicolumn{2}{l}{Three-query ($m$=$3{,}074$)} \\
\midrule
1
& Verify \& Fix & 27.9\%
& Verify \& Fix$\rightarrow$Verify \& Fix & 11.8\%
& Verify \& Fix$\rightarrow$Verify \& Fix$\rightarrow$Verify \& Fix & 5.5\% \\

2
& Implement & 23.5\%
& Implement$\rightarrow$Verify \& Fix & 7.4\%
& Implement$\rightarrow$Verify \& Fix$\rightarrow$Verify \& Fix & 3.1\% \\

3
& Get Unstuck & 21.5\%
& Get Unstuck$\rightarrow$Verify \& Fix & 6.9\%
& Get Unstuck$\rightarrow$Verify \& Fix$\rightarrow$Verify \& Fix & 2.6\% \\

4
& Provide Info & 12.2\%
& Verify \& Fix$\rightarrow$Provide Info & 6.7\%
& Verify \& Fix$\rightarrow$Provide Info$\rightarrow$Verify \& Fix & 2.6\% \\

5
& Understand & 9.5\%
& Get Unstuck$\rightarrow$Provide Info & 6.0\%
& Verify \& Fix$\rightarrow$Verify \& Fix$\rightarrow$Provide Info & 2.4\% \\

6
& Solution Request & 4.0\%
& Implement$\rightarrow$Implement & 5.7\%
& Get Unstuck$\rightarrow$Provide Info$\rightarrow$Provide Info & 2.0\% \\

7
& Peripheral & 1.3\%
& Get Unstuck$\rightarrow$Implement & 4.4\%
& Implement$\rightarrow$Implement$\rightarrow$Verify \& Fix & 2.0\% \\

8
& - & -
& Get Unstuck$\rightarrow$Get Unstuck & 4.4\%
& Implement$\rightarrow$Implement$\rightarrow$Implement & 2.0\% \\

9
& - & -
& Implement$\rightarrow$Provide Info & 3.7\%
& Verify \& Fix$\rightarrow$Provide Info$\rightarrow$Provide Info & 1.9\% \\

10
& - & - 
& Verify \& Fix$\rightarrow$Implement & 3.6\%
& Get Unstuck$\rightarrow$Provide Info$\rightarrow$Verify \& Fix & 1.8\% \\

\bottomrule
\end{tabular}
\caption{\textbf{Most Common Queries in Short Conversations} 
--
\textmd{{\small  
Top sequences for conversations with $\leq$3 queries, by count. Percentages are relative to all conversations of each length, and the listed patterns cover 100.0\% of single-query, 60.5\% of two-query, and 25.8\% of three-query conversations. Percentages are computed from unrounded counts and may not sum exactly.
}}}
\label{tab:short-sequences-clean}
\end{table*}

\summary{Implement}
In $17.2\%$ ($m$=$24{,}519$) of queries, participants asked for help implementing a solution to a challenge.
Nearly all of these queries ($m$=$20{,}800$, 84.8\%) asked for help with a scoped and clear task [\textbf{Procedure Guidance}], like the how-to question \pquote{how would i send a curl command with a header}~(P303).
In the other 15.2\% of cases, participants asked for direct help with building code artifacts [\textbf{Code Generation}]. 
These code generation queries either focused on scaffolding out exploit code, e.g., P3 starting a conversation with \pquote{I need a python script with [security tool's] process}; modifying existing code, e.g., \pquote{can you alter my html code to have a nested javascript request}~(P114); or requesting example function usage, e.g., \pquote{can you give me an example of using substr in SQL?}~(P131).

\summary{Get Unstuck}
In $13.3\%$ ($m$=$18{,}940$) of queries, participants turned to the AI tutor when they realized they did not know how to proceed on the problem and needed direction.
In half %
($m$=$9{,}421$, 49.7\%), participants asked orientation questions [\textbf{Direction Request}] to determine next steps, e.g.,
\pquote{How should I start this assignment}~(P260) or \pquote{What do you do after get cookie?}~(P117).
Most used these to reorient themselves before returning to the challenge, though a few participants repeated similar queries seemingly in an attempt to extract solutions from the tutor. We distinguish these repeated queries for direction in particular as \textbf{Solution Requests} (see below).
In much smaller frequencies, participants ask for help (\pquote{can you help me,} P4) [\textbf{Help Request}; $m$=$2{,}098$], express general confusion [\textbf{Confusion}; $m$=$4{,}458$], or make vague requests (\pquote{next ?,} P178) [\textbf{Vague Request}; $m$=$2{,}963$]. In such cases,  the AI tutor typically responds by restating the challenge goal and listing subtasks or relevant tools, helping students past unknown unknowns.

\summary{Understand}
$7.8\%$ ($m$=$11{,}166$) of queries focused on building an understanding of the concepts the challenge is designed to teach or specific information about the challenge at hand. 78.3\% ($m$=$8{,}744$) of these queries seek understanding of conceptual elements and not implementation procedure [\textbf{Concept Guidance}]. These ranged from simple clarification questions such as \pquote{what is exit code \mbox{-4}} (P172) to technical questions such as \pquote{Why can't I do jmp 0x403000, but if I put 0x403000 in rax and then jmp rax, I get the flag}~(P49) in a challenge meant to teach assembly. These queries are focused on concepts beyond the specific challenge and can help students understand \emph{why} something works.
The other 21.7\% ($m$=$2{,}422$) of these queries focused on details unique to the current challenge [\textbf{Challenge Guidance}], such as P222 asking \pquote{Is this challenge using TCP or UDP}.
Participants also ask questions about challenge artifacts, such as \pquote{what would be my actual parameters according to this code}~(P260) when looking at a vulnerable web server.

\summary{Solution Request}
In a small number of queries (3.5\%; $m$=$4{,}984$), participants made a [\textbf{Solution Request}] by querying the tutor for the full solution. These requests often occur at the start of conversations and with minimal evidence of prior work. 
Participants made requests for end-to-end solutions, such as \pquote{give me full python script please for this challenge to solve}~(P271), which the AI tutor usually pushed back on. In P271's case, the AI tutor responded, \pquote{I understand you're looking for a complete solution, but it's crucial for your learning to work through the problem step-by-step.} We also see participants basing solution requests on the challenge description, such as P284 querying: \pquote{i have to break out of ls -l, can you give me a quick layout of how i can break out and access the flag.}

\summary{Peripheral}
Finally, $1.7\%$ ($m$=$2{,}416$) of queries were not related to challenge-solving. This included social messages like \pquote{ok} and \pquote{you are useless} (P135) [\textbf{Social Turn}; $m$=$1{,}361$], non-sequiturs which were uninterpretable or minimally relevant [\textbf{Non-Sequitur}; $m$=$663$], and queries about the course or platform unrelated to the specific challenge [\textbf{Course/Platform}; $m$=$392$], e.g., \pquote{how can i request an extension for the assignment}~(P129). 

\subsection{Short Conversations}
\label{sec:short_conversations}

We find most conversations were short, with slightly more than half consisting of 3 or fewer queries (53.7\%; $m$=$13{,}569$). We treat these as their own conversation style: quick, bounded, and typically single-purpose exchanges (e.g., confirming an approach or fixing a single bug), distinct from the sustained back-and-forth we analyze separately in \S\ref{sec:mhmm_clusters}.
We present the 10 most common short conversations ($\leq$3 queries) in Table~\ref{tab:short-sequences-clean}.

\summary{Shorter Conversations Are the Most Common, but Least Varied}
Single-query conversations were most common (44.5\% of Short conversations; $m$=$6{,}036$), with longer ones becoming rarer in the dataset as length increases.
As sequence length increased, patterns inevitably became more fragmented: two-query sequences had 49 unique patterns with the top 5 covering 38.7\%, while three-query sequences had 266 patterns with the top 5 covering only 16.2\%.

\summary{Short Conversations Are Dominated by Debugging, Implementation, and Problem Orientation}
Short conversations are primarily composed of three query types: \textit{Verify \& Fix}, \textit{Implement}, and \textit{Get Unstuck}.
They account for 5 of the 10 most common three-query conversations (15.2\%), 7 of the 10 most common two-query conversations (44.2\%), and they were the top 3 most common single-query conversations (72.9\%).

\summary{Most Short Conversations Are Formed by Common Query Patterns}
Short conversations hold several recurring patterns.

\textit{Starting Conversations With Direction:}
While \textit{Get Unstuck} is a fairly common query, among the top 10 short conversations it appears in only two positions: when beginning a conversation, or as a follow-up to another \textit{Get Unstuck} query.
These sequences make up 4 of the top 10 two-query conversations (21.7\%) and 3 of the top 10 three-query conversations (6.4\%).
Afterward, some participants attempt the challenge and hit errors, others paste challenge materials for the tutor to interpret, and others ask for implementation. 

\textit{Building Up a Solution:}
Participants repeatedly build up a solution via multiple \textit{Implement} queries, or \textit{Implement} followed by \textit{Verify \& Fix} or \textit{Provide Info}.
These make up 3 of the top 10 two-query conversations (16.8\%) and 3 of the top 10 three-query conversations (7.1\%). 
For example, P289 in an AES challenge asked the following sequence of queries: \pquote{do you need to pad this challenge?}$\rightarrow$\pquote{how can I turn the hex of hKey to bytes}$\rightarrow$\pquote{how can I determine what the IV is?}, with each question digging one layer deeper into the challenge.

\textit{Repeated Debugging:}
In an effort to get a working solution, participants often use the AI tutor to repetitively debug code via consecutive \textit{Verify \& Fix}$\rightarrow$\textit{Verify \& Fix} queries, provide additional information to a request via \textit{Verify \& Fix}$\rightarrow$\textit{Provide Info}, or fix an issue and then ask for an example implementation via \textit{Verify \& Fix}$\rightarrow$\textit{Implement}.
These account for 3 of the top 10 two-query conversations (22.1\%) and 6 of the top 10 three-query conversations (18.1\%). Notably, repetitive \textit{Verify \& Fix} is also the most common two- and three-query conversation.
While some participants appear to stall on the same issue, these queries can also show progression through the multiple steps a problem has. For instance, P277 repeatedly asked for help when solving multiple issues in their assembly, shifting from \pquote{What might be causing my program to crash?} in early requests to \pquote{My code is now no longer crashing, but it is not properly lower-casing the strings} in later requests.

\subsection{Long Conversations}
\label{sec:mhmm_clusters}
\begin{figure*}[!t]
  \centering
  \includegraphics[width=\linewidth]{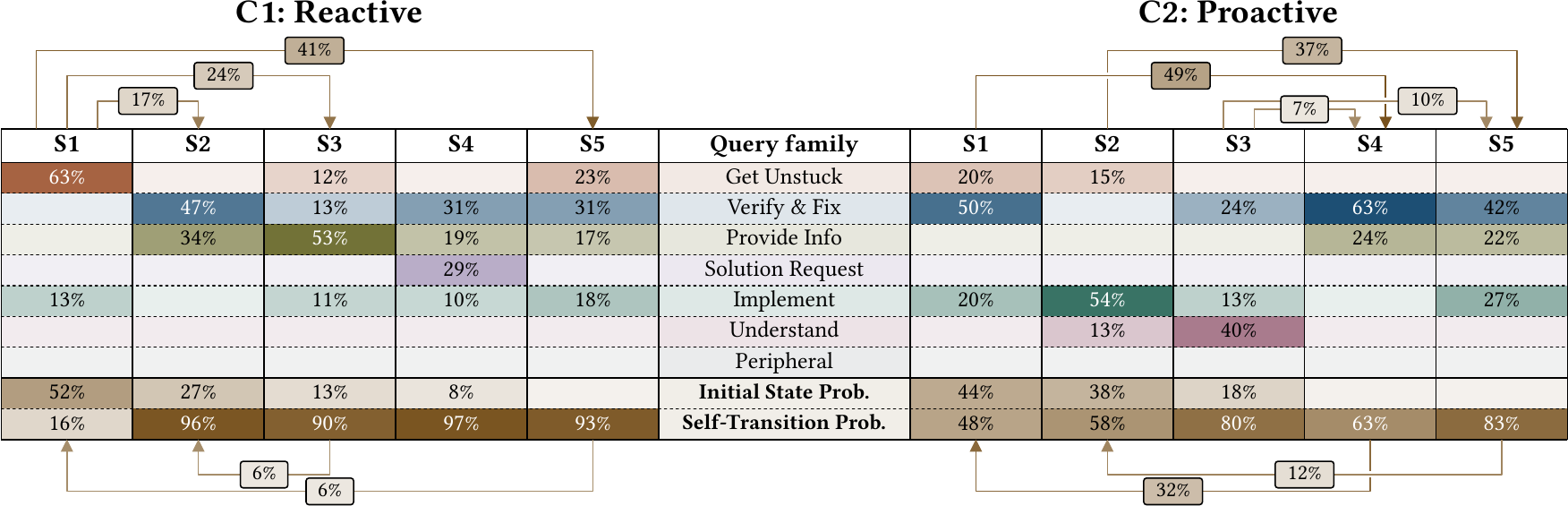}
  \Description{State-transition diagrams of the two fitted conversation clusters, Reactive and Proactive, each with five hidden states, showing emission probabilities for the seven query families and transition probabilities between states.}
  \caption{\textbf{Observed Conversation Styles} 
--
\textmd{{\small  
The conversation structure for $m$=$11{,}692$ query sequences. Columns are hidden states (S1--S5), rows are emission probabilities per query family, and the bottom two rows give initial-state and self-transition probabilities. Arrows show only transitions $\geq$5\% and states show only emissions $\geq$10\%.
}} 
}
  \label{fig:mhmm_cluster}
\end{figure*}

To understand longer conversational patterns, we analyze conversations of 4 queries or longer ($m$=$11{,}692$; 46.3\%).
Similar to prior work modeling complex patterns for problem-solving~\cite{ouyang2023artificial},
we use an MHMM to identify patterns.

We use MHMMs as they are particularly well-suited for modeling sequences in which
(1) sequences are of varied length and
(2) observed outputs map to hidden states~\cite{rabiner1989tutorial}.
In our case, the \textit{states} represent a student's goals at a given point in the conversation; in a particular state, a student is more likely to send queries of a particular type. 
A \textit{cluster} represents a conversation style: a set of states (student goals) and transitions characterizing how students' goals shift during the conversation.
Following best practices, our MHMM was fit to 11,692 conversations with at least 4 queries from 303 students (truncated to 22 queries in length~\cite{jinglin2019}),
and selected according to the Integrated Completed Likelihood criterion~\cite{biernacki2000icl}.

\summary{Long Conversations Are Reactive or Proactive}
As shown in Figure~\ref{fig:mhmm_cluster}, our fitted MHMM yielded 2 clusters of 5 states, capturing two distinct conversation styles which we denote as ``Reactive'' and ``Proactive'', named based on their states and structural makeup.
Broadly, Reactive styles account for 51.4\% ($m$=$6{,}008$) of conversations; in these, participants provide context and ask the AI tutor to identify problems.
The remaining 48.6\% of conversations are Proactive, in which participants first produce their own solution to some extent and then ask the tutor to verify it, answer questions about specific subtasks of the challenge, or address targeted questions that help the student better understand the underlying concepts.

\summary{Reactive Conversations Begin With General Confusion, Proactive With Focused Questions}
As the Initial State Prob. row of Figure~\ref{fig:mhmm_cluster} shows, the initial-state probabilities differed between Reactive and Proactive conversations. Notably, over half of Reactive conversations (52\%) enter through a single state (S1)  which was dominated by \textit{Get Unstuck} queries, and 27\% enter through a state focused on providing context and fixing bugs. This suggests Reactive conversations arrive either directionless or focused on fixing a bug.

Conversely, Proactive conversations enter primarily through two states: S1 (44\%) and S2 (38\%). 
While both feature some \textit{Get Unstuck} queries to help with challenge orientation, bug-fixing and code implementation dominate S1, and code implementation dominates S2 with some focus on understanding.
This may imply Proactive conversations begin with a targeted entry point rather than general disorientation, entering mid-debugging or with implementation tasks and concept questions.

\summary{Reactive Conversations Converge, Proactive Cycle}
After starting at an initial state, the conversation may transition to other states; the probability of this is shown via the arrows and the ``self-loop'' probability in Figure~\ref{fig:mhmm_cluster}.
Together, these indicate the likelihood that the next query will remain in the same state and determine whether a state is transient (quickly exited), absorbing (likely to stay in the state), or a hub (cycled through repeatedly).

Reactive conversations quickly settle into one of four absorbing states: verification paired with context provision, context-dumping loops, confused-verification cycling, or verification with solution requests. 
For example, while working on a firewall challenge, P205 ended on a nine-query loop containing: \pquote{REJECT?}, \pquote{it didnt do anytihng}, \pquote{this is what im using and its not doing anyting}, \pquote{so did not work}, and pasted \texttt{iptables} commands. This oscillation showcases the absorbing verification/context loop.
Proactive conversations more often cycle between states as students work through the challenge, potentially building on what came before.
Conversations begin in three states with three different focuses: verification, implementation, or understanding.
Three hub states (S1, S3, and S4) cycle among themselves: for example, students move from verification to context-checking and back (S1$\rightarrow$S4 at 49\%, S4$\rightarrow$S1 at 32\%).

\summary{Reactive Conversations Provide Info Ubiquitously, Proactive Selectively}
Looking at Figure~\ref{fig:mhmm_cluster}, we see \textit{Provide Info} queries in all four absorbing states (S2--S5) in the Reactive cluster, while they only occur at notable rates in two states (S4 and S5) in Proactive conversations, indicating that participants in Reactive conversations utilize \textit{Provide Info} queries more often compared to Proactive ones.
Reactive conversations instead treat the tutor as an oracle: participants submit artifacts and wait for explanation.
In S2 we see this as an absorbing verification-context loop, such as P194 submitting a seven-query sequence of pasted code, error messages, and \pquote{how does this look?} queries, without directly interpreting any of the tutor's responses.
This mirrors the short-sequence evidence-provision pattern (Verify \& Fix$\rightarrow$Provide~Info), but Short conversations resolve after a few turns while the Reactive absorbing topology sustains these loops for more exchanges.

Proactive conversations also contain context queries, but the context serves a different purpose. Rather than standing alone as an artifact for the tutor to interpret, context tends to co-occur with targeted queries. For example, P199 on a reverse-engineering challenge needed to craft a valid input for a parser, but rather than simply dumping code, P199 provided context and asked targeted questions: \pquote{how can I find the pixel values I need?} and \pquote{if I use binary ninja where can I find the value I need to solve the challenge?}, supplemented with decompiled C code to understand the required format.
Here, each piece of context is evidence for a specific question or problem the student is working to overcome.

\summary{Reactive Conversations Ask for Solutions}
A distinctive minority of Reactive conversations included explicit attempts to outsource problem-solving to the AI tutor. 
While both styles contain \textit{Solution Request} queries, only the Reactive cluster contains a dedicated absorbing state where 29\% of emissions are solution requests. 
Overall, 25.8\% of Reactive conversations contain at least one \textit{Solution Request} versus 11.8\% of Proactive conversations, and 10.9\% versus 1.8\% contain more than one.
When the tutor does not provide solutions, participants ask again, often with frustration.
For instance, to complete a cryptography challenge, P113 sent six solution-request queries: \pquote{provide me the commands of the challenge step to step} and repeated it with minor variations until the tutor eventually provided enough information for a solution.
When the AI tutor pushes back on solution requests, it asks students what they have tried and learned. P249, in one such instance, responded \pquote{yes i tried} and followed with \pquote{give correct commands pleaseee}.
Lastly, we see participants pivot to solution requests after extended struggling, such as P133 asking the AI tutor~\pquote{can you just give me the answer?} after 13 queries debugging their x86-64 code.
This pattern is largely absent from Proactive conversations. 

\summary{Proactive Conversations Ask for Insights}
The Proactive cluster uniquely contained a state for conceptual inquiry (40\% understanding, 24\% verification) that participants cycled through to build understanding.
For example, P117 on an assembly challenge asked three consecutive \textit{Understand} queries to construct a mental model of the stack: \pquote{could you explain how stack and base pointer are being accessed in this line of code?}, then \pquote{what are the differences between the stack pointer and base pointer?}, then \pquote{What is the impact of me reading the address of base pointer when input buffer initialized versus before read() being called?} 
However, this sustained \textit{Understand} query style is relatively rare; more often, \textit{Understand} queries appear within implementation-heavy conversations to check understanding, e.g.,
\pquote{so the correct password isn't 1010 but rather it is in bytes? and my goal is to convert it to said bytes?}~(P28).
This allowed P28 to verify their mental model of the problem and continue.

\summary{Reactive Debugging Is AI-Driven, Proactive Debugging Is Student-Driven}
While debugging is common across Reactive and Proactive conversations, they differ in how they relate to other potential states.
In Reactive conversations, states with high debugging probabilities (S2, S4, and S5) are often not left, with self-looping transitions $\geq$93\%. %
This may manifest as repetitive trial-and-error queries where participants repeatedly rely on the tutor to identify the problem rather than acting on the response before submitting another query.
For example, P173 asked, \pquote{can you check input.py and see if my code is correct}. After the tutor suggested a fix, P173 asked for help implementing, \pquote{how do I directly convert from binary to hex}, then \pquote{can you check input.py now}. When the tutor confirmed P173's overall direction, they replied \pquote{its not working} and again asked, \pquote{how do I encode using latin-1,} then \pquote{its not working} again. 

In contrast, Proactive states with high debugging probabilities (S1, S4, and S5) are more likely to cycle between understanding and implementation-focused states like S2.
This takes the form of specific debugging queries that participants then use to build a solution through \textit{Implement} or \textit{Understand} queries.
For example, P198 was working on a base64 challenge and queried \pquote{I have this for the entered password so far [...] it matches that, but it is still incorrect.} After two debugging exchanges, the student identified the root cause as a newline character, \pquote{is it because there is an endline?,} and immediately pivoted to implementation in an attempt to fix the problem: \pquote{how would I load this into a file to pipe into the challenge}.

\begin{table}[t]
\setlength{\tabcolsep}{3pt}
\renewcommand{\arraystretch}{1.0}
\centering
\small
\begin{tabular}{@{}l r r r@{}}
\toprule
Module & Short & Reactive & Proactive \\
\midrule
Linux          & 66.4\% (2{,}524) & 21.8\% (827)  & 11.9\% (451) \\
Data \& Access & 48.9\% (1{,}482) & 28.3\% (856)  & 22.8\% (691) \\
Web \& SQL     & 60.4\% (2{,}356) & 19.5\% (760)  & 20.1\% (783) \\
Web Sec.       & 38.2\% (1{,}075) & 33.8\% (951)  & 28.0\% (789) \\
Comp.\ Arch.   & 58.8\% (2{,}425) & 18.2\% (752)  & 23.0\% (947) \\
Network Sec.   & 45.8\% (1{,}421) & 26.7\% (828)  & 27.5\% (854) \\
Cryptography   & 48.4\% (1{,}032) & 23.4\% (500)  & 28.2\% (602) \\
Reverse Engr.  & 56.3\% (807)     & 21.3\% (305)  & 22.4\% (321) \\
Binary Sec.    & 48.5\% (447)     & 24.8\% (229)  & 26.7\% (246) \\
\bottomrule
\end{tabular}
\caption{\textbf{Conversation Style Distribution by Module} --
\textmd{{\small We present the distribution of conversations classified as Short, Reactive, or Proactive within each module.}}}
\label{tab:style_distribution}
\end{table}

\summary{Style Distribution Across Modules}
We also find that the distribution of conversation styles
varied across modules.
Among long conversations, the distribution of Reactive vs. Proactive conversations varied significantly by module ($\chi^2(8) = 168.04$, $p < .001$; Cram\'er's $V = .120$). 
Table~\ref{tab:style_distribution} reports the three-way breakdown by module. Short conversations were the most common style in every module but their share varied substantially (38.2\%--66.4\%), while Reactive conversations ranged from 18.2\% to 33.8\% and Proactive from 11.9\% to 28.2\%.
The Linux and Web \& SQL modules were dominated by Short conversations (66.4\% and 60.4\%, respectively), consistent with their simple structure where just one or two commands can solve the challenge.
Conversely, Web Security had the lowest Short conversation share (38.2\%) and the highest Reactive share (33.8\%); its multi-step exploitation workflows (crafting payloads, observing outcomes, adjusting) naturally extend conversations beyond two or three queries.
Among long conversations specifically, the Reactive share declined from 64.7\% in the introductory module to roughly 44--49\% from module~5 onward, suggesting students shifted toward Proactive conversations as they gained experience with the tutor; however, total conversation volume also declined across the semester (from 3{,}802 down to 922), so this shift may partly reflect survivorship bias rather than individual learning.

\begin{figure*}[t]
  \centering
  \includegraphics[width=\linewidth]{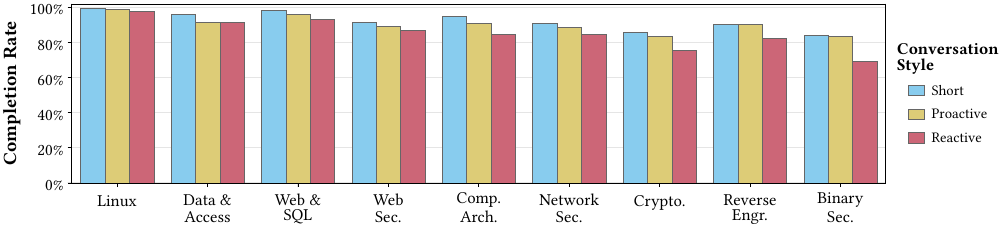}
  \Description{Bar chart of completion rate by module for Short, Proactive, and Reactive conversations. Short bars are highest in nearly every module and Reactive lowest, with the gap widening in later modules.}
  \caption{\textbf{Module Completion by Conversation Style} -- \textmd{\small Observed completion rates by module and style.}}
  \label{fig:completion-by-module}
\end{figure*}

\section{Performance Across Conversations (\ref{rq:performance})}
\label{ref:performance}
Having identified three conversation styles (\S\ref{sec:short_conversations} and \S\ref{sec:mhmm_clusters}), we examine how conversation styles relate to challenge completion.

\summary{Summary Statistics}
By dividing the participants' 25,261 conversations across the style categories of Short (53.7\%, $m$=$13{,}569$), Reactive (23.8\%, $m$=$6{,}008$), and Proactive (22.5\%, $m$=$5{,}684$), we aggregate their respective completion rates across modules in Figure~\ref{fig:completion-by-module}.

Looking across styles, we find that Short conversations had the highest overall completion rate (94.6\%), followed by Proactive (90.7\%) and Reactive (87.5\%). This ordering held in every module except Reverse Engineering, where Proactive conversations narrowly exceeded Short.
Short conversations' consistently high completion rates likely reflect participants arriving with a specific question rather than an unresolved problem, not that brief interaction is itself more effective.
Across modules, we find completion rates declined throughout the semester, from 99\% in Linux to 80.4\% in Binary Security, while the gap between Proactive and Reactive conversations widened from less than 1\% to 14.3\%.

\begin{table}[t]
\setlength{\tabcolsep}{3pt}
\renewcommand{\arraystretch}{1.0}
\centering
{
\small
\begin{tabular}{@{}l r r r@{}}
\toprule
\textbf{Factor} & \textbf{LR $\chi^2$} & \textbf{df} & \textbf{$p$-value}\\
\midrule
\multicolumn{4}{l}{\textit{Primary Effects}} \\
\hspace{3mm}{Module} & \textbf{965.27} & 8 & \textbf{$<$.001} \\
\hspace{3mm}{Conversation Style} & \textbf{168.18} & 2 & \textbf{$<$.001} \\
\multicolumn{4}{l}{\textit{Interaction}} \\
\hspace{3mm}{Module $\times$ Conversation} & \textbf{46.67} & 16 & \textbf{$<$.001} \\
\bottomrule
\end{tabular}
}
\caption{\textbf{Factors Associated With Challenge Completion} --
\textmd{\small Via likelihood ratio tests comparing our fitted mixed-effects model to one without each fixed effect, we find that all fixed effects are significantly associated with challenge completion.
\textbf{Bold} indicates $p < .05$.}
}
\label{table:analysis-of-deviance}
\end{table}

\subsection{Statistical Analysis}
\label{sec:statistical_analysis}

Given the repeated-measures structure of our data, where the same participants contribute conversations across challenges, we used a mixed-effects logistic regression to model this data appropriately. We modeled challenge completion as a function of conversation style, module, and the interaction effect between conversation style and module (i.e., conversation styles may have different effects depending on the module). To account for participant-specific differences in performance, we modeled participants as a random effect.

\summary{Conversation Style and Module Are Associated With Completion}
To evaluate whether each fixed effect is significantly associated with challenge completion, we conduct Type II likelihood ratio tests~\cite{winter2013linear}. %
As shown in Table~\ref{table:analysis-of-deviance}, we find module ($\chi^2(8) = 965.27$, $p < .001$) and conversation style ($\chi^2(2) = 168.18$, $p < .001$) significantly predict challenge completion (\ref{rq:performance}). 
We also find a significant interaction between these effects ($\chi^2(16) = 46.67$, $p < .001$), indicating that the association between style and completion varies by module.
To avoid misleading conclusions, we evaluate effects within each module rather than interpret main effects independently~\cite[pp.~164--169]{purchase2012experimental}.

\begin{table}
\setlength{\tabcolsep}{3pt}
\renewcommand{\arraystretch}{1.0}
\centering
\footnotesize
\begin{tabular}{@{}l r r r r r r r@{}}
\toprule
& & \multicolumn{2}{c}{\footnotesize Short--React} & \multicolumn{2}{c}{\footnotesize Short--Proact} & \multicolumn{2}{c}{\footnotesize Proact--React} \\
\cmidrule(lr){3-4} \cmidrule(lr){5-6} \cmidrule(lr){7-8}
Module & $m$ & OR & $p$ & OR & $p$ & OR & $p$ \\
\midrule
Linux
& 3,802
& \textbf{3.64} & \textbf{.002}
& 2.37 & .249
& 1.54 & .695 \\

Data \& Access
& 3,029
& \textbf{2.29} & \textbf{$<$.001}
& \textbf{2.56} & \textbf{$<$.001}
& 0.89 & .829 \\

Web \& SQL
& 3,899
& \textbf{3.74} & \textbf{$<$.001}
& \textbf{2.76} & \textbf{$<$.001}
& 1.35 & .422 \\

Web Sec.
& 2,815
& \textbf{1.64} & \textbf{.004}
& \textbf{1.57} & \textbf{.022}
& 1.05 & .957 \\

Comp. Arch.
& 4,124
& \textbf{3.61} & \textbf{$<$.001}
& \textbf{2.49} & \textbf{$<$.001}
& \underline{1.45} & \underline{.058} \\

Network Sec.
& 3,103
& \textbf{1.64} & \textbf{.001}
& \textbf{1.51} & \textbf{.017}
& 1.09 & .854 \\

Cryptography
& 2,134
& \textbf{1.75} & \textbf{$<$.001}
& \textbf{1.64} & \textbf{.004}
& 1.07 & .908 \\

Reverse Engr.
& 1,433
& 1.44 & .170
& 0.92 & .927
& 1.57 & .183 \\

Binary Sec.
& 922
& \textbf{1.92} & \textbf{.005}
& 1.19 & .728
& 1.62 & .113 \\

\bottomrule
\end{tabular}
\caption{\textbf{Association Between Conversation Styles and Completion Across Modules} --
\textmd{{\small We report the estimated associations between conversation styles and the likelihood of challenge completion.
OR$>$1 indicates that AI tutor conversations of the first style are more likely to result in a completed challenge than conversations of the second style.
\textbf{Bold} indicates $p < .05$; \underline{underline} indicates $p < .10$.}}}

\label{tab:per_module_contrasts}
\end{table}

\summary{Reactive Conversations Correlate With Less Completion Than Short Conversations Across Most Modules}
After performing separate Tukey-adjusted post-hoc pairwise tests between the different conversation styles for each module, we find
that in eight of nine modules, Reactive conversations are associated with significantly lower completion rates than Short conversations (Table~\ref{tab:per_module_contrasts}). 
Across these eight modules, odds ratios (ORs) ranged from 1.64 to 3.74 (a 39--73\% decrease in the odds of completion for Reactive conversations).
The Reactive--Short gap varied across modules: it was more pronounced in modules such as Computer Architecture and Binary Security,  where failed attempts often produce minimal error messages that give little insight and do not identify the root cause of the failure.
Because Reactive students rely on pasting their current state and asking the tutor for interpretation, an uninformative failure leaves the tutor little to work from, and it can only respond with generic suggestions or guesses.
By contrast, the gap is smaller in modules such as Web Security or Network Security, where students craft a payload, observe its outcome, and adjust, so attempts can yield partial progress.
Reverse Engineering was the one module where Reactive did not differ significantly from Short, and the only module in which completion rates for participants with Proactive conversations closely matched those with Short conversations (90.7\% vs.\ 90.2\%). Because the module focuses on iteratively constructing inputs to satisfy a parser and provides specific error messages (e.g., ``ERROR: Invalid magic number!'', ``ERROR: Incorrect width!''), both Reactive and Proactive conversations follow similar test-then-fix patterns. This tight feedback loop may explain the small difference between Reactive and Short conversations, since students do not have to guess why their solution fails.

\summary{Proactive Conversations Are Associated With Less Completion Than Short Conversations Across Several Modules}
Proactive conversations also showed significantly lower completion than Short conversations across several modules, with odds ratios ranging from 1.51 to 2.76 (a roughly 34--64\% decrease for Proactive conversations) in six of nine modules.
We did not find any differences for the other three modules. 
The first was the introductory Linux module, where challenges could be solved by running one or two commands; we observed significantly fewer Proactive conversations and all types had high completion rates, likely contributing to the lack of any observed differences.
The second was Reverse Engineering, which as discussed previously has built-in feedback that reduces the difference between conversation styles. Finally, Binary Security also showed no significant difference, likely due to a smaller sample size and course attrition limiting statistical power.

\summary{Pairwise Proactive vs.\ Reactive Contrasts Did Not Reach Significance Within Any Module}
Although Proactive conversations showed equal or higher raw completion rates than Reactive in all nine modules, no per-module contrast achieved statistical significance.
The closest were the Computer Architecture (OR $= 1.45$ [0.99, 2.12], $p = .058$)
and Binary Security (OR $= 1.62$ [0.92, 2.84], $p = .113$) modules.
While this may be because modules like Binary Security are the smallest and may have limited statistical power, we cannot draw any conclusions.

\summary{Neither Student Background nor AI Use Predicts Challenge Completion}
To investigate whether students' background, i.e.,
self-reported prior cybersecurity knowledge (Q1), or frequency of general AI use outside the course (Q2)
predict challenge completion, we added these self-reported metrics as independent variables in our mixed-effects regression described above and re-ran our analysis. We \textit{do not} find evidence that either factor significantly correlates with performance, while the statistically significant associations with module and conversation style remain when controlling for student background (\S\ref{appendix:confounding}).

\begin{figure}[t]
\centering

\includegraphics[width=\linewidth]{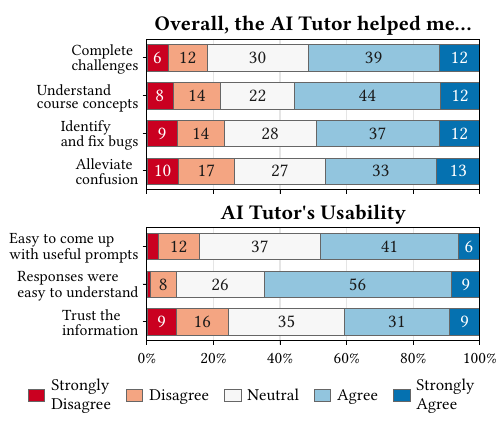}
\Description{Stacked bar chart of Likert ratings for four utility and three usability items.}
\caption{\textbf{Perceived Utility and Usability of the AI Tutor} 
-- \textmd{{\small Participants' Likert ratings of the AI tutor's usefulness (Q4--Q7) and usability (Q10--Q12) across tasks.}}
}
\label{fig:survey-utility}
\end{figure}

\section{Perceptions of the AI Tutor (\ref{rq:student-beliefs})}
\label{sec:perceptions}
We now present students' reported perceptions of our AI tutor via our end-of-semester survey (\S\ref{appendix:eos-survey}) with $n$=$293$ participants.

\subsection{Utility and Usability of the AI Tutor}
\label{sec:perceived_utility}
Participants rated the tutor's utility and usability on a Likert scale across seven items (Figure~\ref{fig:survey-utility}), and answered two open-response questions on which tasks the AI tutor was most or least useful.

\summary{While Useful, the Tutor Is Better at Solving Issues Than Alleviating Confusion}
Roughly half of participants agreed that the tutor helped them understand concepts (55.6\%; $n$=$163$), complete challenges (51.5\%; $n$=$151$), find bugs (49.1\%; $n$=$144$), and alleviate confusion (46.4\%; $n$=$136$).
To evaluate whether these differences between task types are significant, we conducted a Friedman test and found that significant differences existed ($\chi^2(3) = 13.21$, $p = .004$).
Then, via post-hoc pairwise Wilcoxon signed-rank tests with Bonferroni correction, we found that participants agreed that the tutor helped more in completing challenges (W $=$ $3{,}034.5$, $p = .01$) and
understanding concepts (W $=$ $3{,}187.0$, $p = .04$)
than in alleviating confusion.

\summary{Students Valued Direction-Seeking but Were Split on Debugging}
When asked which tasks the AI tutor was most and least useful for (Q8,
Q9), 
\emph{direction-seeking and reconnaissance} was the most frequently cited task ($n$=$169$) and predominantly positive (119 pos. vs.\ 50 neg.), with P47 finding the tutor useful for \pquote{getting some sort of direction in solving the problems\ldots a good initial idea to work off of.} 
\emph{Concept guidance} ($n$=$53$) and \emph{code generation} ($n$=$49$) were less frequently cited but also largely positive (39 vs.\ 14 and 38 vs.\ 11, respectively), with P120 stating that it was helpful for \pquote{conceptual questions and [they] would get a useful response[s]\ldots it was great at breaking down lines of code, if I saw a new unfamiliar function[s] or one I had forgotten.} 
\emph{Debugging} ($n$=$103$) drew the most mixed assessments, with slightly more negative than positive mentions (54 neg. vs.\ 49 pos.) and participants praising the tutor's ability to catch syntactic errors but finding it unhelpful for deeper reasoning.

\label{sec:perceived_usability}

\summary{Responses Were Easier to Read Than to Prompt For or Trust}
Most found responses easy to understand (64.8\%; $n$=$190$), and roughly half agreed it was easy to create useful prompts (47.8\%; $n$=$140$).
However, trust was notably lower: only 40.6\% ($n$=$119$) agreed they could trust the tutor's information.
A Friedman test confirmed significant differences ($\chi^2(2) = 46.29$, $p < .001$). Post-hoc tests showed responses were easier to understand than to prompt for (W $=$ $2{,}464.0$, $p < .001$) or to trust (W $=$ $2{,}960.0$, $p < .001$), and that prompting was rated higher than trust (W $=$ $5{,}216.0$, $p = .005$).

\summary{Unfavorable Ratings Reflect Tutor Limitations, Not Ineffective Use}
If unfavorable ratings reflected ineffective use rather than the tutor's limitations, participants with more Proactive conversations should rate the tutor more favorably.
We correlated each participant's Proactive share (Proactive / (Proactive + Reactive) among their long conversations) with their utility and usability ratings (Q4--Q7, Q10--Q12; Spearman, Holm-corrected; $n$=$289$, as 4 of the 293 participants had only Short conversations and thus no defined share).
No individual rating item was associated with Proactive share after correction except trust: all utility items had Holm-adjusted $p \geq .50$ (Q4--Q7 composite: $\rho = -.10$, $p = .10$, uncorrected), while trust (Q12) significantly decreased with Proactive use ($\rho = -.23$, Holm-adjusted $p < .001$).
Unfavorable ratings therefore appear to reflect the tutor's limitations rather than ineffective use, consistent with its reported decline on harder modules (\S\ref{sec:discontinuation}).

\subsection{Comparison to Human TAs}
\begin{figure}[t]
\centering

\includegraphics[width=\linewidth]{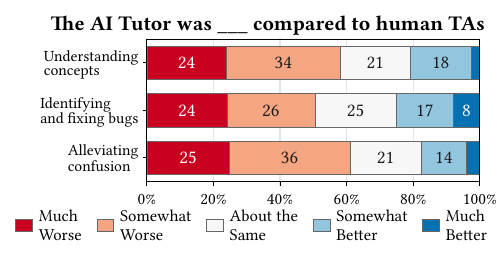}
\Description{Stacked bar chart comparing the tutor to human TAs on three tasks. Most students rate the tutor the same as or worse than TAs, most negatively for alleviating confusion.}
\caption{\textbf{AI Tutor vs.\ TA Comparison}
-- \textmd{\small Participants' ratings comparing the AI tutor to human TAs (Q14--Q16).}}%
\label{fig:survey-comparion-to-ta}
\end{figure}

Participants rated the AI tutor against human TAs on a Likert scale across three items (Figure~\ref{fig:survey-comparion-to-ta}) and answered one open-response question comparing the two.

\summary{Students Preferred TAs for Quality but Valued the Tutor's Availability}
Participants who had not used the course TAs for the specified use case could answer ``N/A'' to these items (e.g., if a participant had not used a TA to help them debug), so the number of comparisons varies by question (Figure~\ref{fig:survey-comparion-to-ta}).
Among those who did provide a comparison,
a majority rated the tutor worse across all dimensions: alleviating confusion (61.1\%; 138 of 226), understanding concepts (58.1\%; 136 of 234), and finding bugs (50.4\%; 113 of 224).
Among the 217 participants who rated all three items, a Friedman test confirmed significant differences across them ($\chi^2(2) = 8.23$, $p = .016$). Post-hoc corrected tests revealed that bug-fixing was rated significantly better than both concept understanding (W $=$ $2{,}515.5$, $p = .021$) and confusion alleviation (W $=$ $1{,}765.5$, $p = .011$), while the latter two did not differ significantly (W $=$ $2{,}186.5$, $p = .58$).
In contrast, in the free-response TA comparison (Q17), the most commonly cited advantages of the tutor were \emph{availability} ($n$=$25$) and \emph{lower social barriers} ($n$=$8$).
P48 noted \pquote{I can use [AI tutor] whenever I want\ldots that's the most dominant advantage,} and P79 highlighted that they didn't feel judged by the AI tutor.

\subsection{Reasons for Stopping AI Tutor Use}
\label{sec:discontinuation}
In Q13, 168 (57.3\%) participants reported stopping or reducing their use of the tutor during the semester; $157$ of these provided a codable reason, which we now evaluate.

\summary{Difficulty With Harder Challenges}
The most frequently cited reason for stopping use was that the tutor was \emph{bad at harder challenges} ($n$=$77$; 45.8\% of those who reported stopping), reinforcing participants' reports that the tutor was useful early but less so on harder material.
Participants consistently reported the tutor's usefulness dropping sharply across modules 5--7, corresponding to the modules where the Reactive--Short completion gap widens substantially (Figure~\ref{fig:completion-by-module}).
Further, we observe a decrease in conversations from module 5 onward (Table~\ref{tab:module-usage}).
P20 noted, \pquote{around cryptography, I quickly found out [AI tutor] was going to be of no help,} and P36 reported, \pquote{the modules just became too complicated.}

The difficulties participants cited were often not generic complaints but mapped to module characteristics. For example, modules with opaque errors such as Computer Architecture or Binary Security made it difficult for students to observe partial progress. Many participants complained of generic debugging advice, e.g., \pquote{stopped [using AI tutor] once we got to working with the stack/IDA as it just repeated steps that I found confusing} (P289). In a conceptually driven module like Cryptography, participants reported trouble translating the AI tutor's explanations into implementation, such as P114 finding the tutor \pquote{useful in understanding what each encryption did} but \pquote{failing to help users construct a successful script caused me to stop using [AI tutor]}. This pattern of opaque errors yielding generic guidance surfaced often and aligns with our quantitative findings that the completion gaps between styles widen the most in these modules. 

\summary{Poor Responses}
\emph{Repetitive responses} ($n$=$28$), \emph{wrong or misleading information} ($n$=$20$), and \emph{vague responses} ($n$=$9$) further drove discontinuation.
Repetitive response complaints describe the student-side observations of the \emph{absorbing states} identified in Reactive conversations (\S\ref{sec:mhmm_clusters}).
Where the MHMM captures these as cycles of sustained verification or context-dumping loops, participants described the same phenomenon: P93 reported the tutor \pquote{just keeps going in the same circle which did not take me anywhere,} and P162 said the tutor would \pquote{copy [and] paste the same thing.} 
Wrong or misleading information ($n$=$20$) compounded the issue: P5 noted the tutor was sometimes \pquote{blatantly incorrect,} and they would have to correct it.

\summary{Preference for Other Resources}
Thirteen participants reported switching to other resources entirely, commonly citing the course Discord and other LLMs: P125 \pquote{became more active in the discord and found it much more useful and constructive,} while P305 \pquote{did switch over to using ChatGPT o3\ldots because it did sometimes seem like [AI tutor] would not really think that hard about the problem.}

\summary{Educational Concerns and Outgrowing the Tutor}
Finally, ten participants expressed \emph{educational concerns}, worrying that relying on the tutor might hinder their own learning.
P26 stated \pquote{at some point I just wanted to understand for myself what to do and [AI tutor] can't do that.}
Similarly, 11 participants reported \emph{outgrowing} the tutor, finding that their skills had advanced beyond what it could offer, such as P78 stating: \pquote{I stopped needing it as much to give me syntax.}

\section{Discussion}
\label{sec:discussion}

In summary, we find that students used the AI tutor heavily, submitting $142{,}526$ queries across seven distinct categories, with the most common being \textit{Verify \& Fix} (checking students' work, 34.7\%), \textit{Provide Info} (supplying context, 21.8\%), and \textit{Implement} (coding a solution, 17.2\%) queries (\ref{rq:interaction}).
Most conversations were short and single-purpose, while longer conversations separated into two styles: Reactive conversations, characterized by context-dumping, absorbing verification loops, and AI-driven debugging, and Proactive conversations, characterized by student-driven problem solving and targeted inquiry that cycles between implementation and understanding (\S\ref{sec:queries}, \S\ref{sec:mhmm_clusters}).
We then examine how these styles relate to challenge completion (\ref{rq:performance}), finding that Short conversations had the highest completion rate (94.6\%), followed by Proactive (90.7\%) and Reactive (87.5\%), and that the gap between Proactive and Reactive conversations widened as material grew harder, from under 1\% in the first module to 14.3\% in the most advanced (\S\ref{sec:statistical_analysis}).
Finally, we asked students for their opinions on the AI tutor (\ref{rq:student-beliefs}), finding that roughly half judged it useful, that a majority rated it worse than human TAs on every dimension measured, and that 57.3\% stopped or reduced their usage during the semester.
From these results, we consider how these conversation styles relate to established help-seeking behavior, provide recommendations for educators deploying AI tutors, and discuss implications for the design of future systems.

\subsection{Conversation Styles in Help-Seeking Theory}
Our identified Reactive and Proactive conversation styles map onto established help-seeking profiles~\cite{karabenick2018academic}:
Reactive conversations
resemble \textit{executive} help-seeking, where students seek assistance to complete tasks without deeply engaging with the underlying material, and
Proactive conversations
resemble \textit{adaptive} help-seeking, where students build understanding through deliberate, focused requests.
This is consistent with the longstanding finding that question quality, not frequency, correlates with achievement~\cite{graesser1994question}, and with traditional CS domains where
unproductive help-seeking behaviors such as hint abuse predict poorer performance~\cite{jost2024impact, mcnichols2025studychat, marwan2020unproductive}.
Our work extends these findings to hands-on cybersecurity education, where the open-ended and cross-domain nature of the challenges makes the distinction between adaptive and executive help-seeking even more important.
Because these are known student behaviors rather than something our tutor created, they are both teachable and detectable.

\subsection{Recommendations for Educators}
\summary{Deploy AI Tutors to Complement, Not Replace, TAs}
Students preferred human TAs across every measured dimension and often complained that the AI tutor's usefulness declined on harder challenges, though they found it useful for scaffolding and syntactic debugging (\S\ref{sec:perceptions}).
Notably, these unfavorable assessments were not confined to students using the tutor ineffectively: ratings did not improve with more Proactive use (\S\ref{sec:perceived_utility}), suggesting they reflect genuine tutor limitations.
Students also rarely asked the tutor for conceptual help: conceptual queries made up only 7.8\% of total volume, and unlike a human TA who may organically weave conceptual explanation into debugging help, the AI often only did so when students explicitly asked.
We therefore recommend deploying AI tutors as a complement to human help, and framing the tutor as a resource for problem orientation and debugging, while encouraging students to seek human help for conceptual depth.

\summary{Educate Students on How to Use AI Tutors}
Students shifted between Reactive and Proactive conversations across the semester, and Proactive conversations showed equal or higher raw completion rates than Reactive in every module. No single module reached significance (\S\ref{sec:statistical_analysis}), though the average across modules did (\S\ref{appendix:extended-analysis}). Encouraging Proactive engagement therefore remains a defensible, if modest, recommendation.
We also did not observe systematic prompt engineering attempts to circumvent Socratic guardrails. The few solution-seeking attempts consisted of repetition under frustration, suggesting light query auditing plus Socratic guardrails could be enough in practice.
Educator effort therefore could be better directed at metacognitive feedback~\cite{ROLL2011helpseeking} such as self-assessment, structured reflections, or requiring more detailed reasoning from the student when requesting help, thus nudging students to more Proactive approaches.
Educators can also build clearer feedback into their challenges, since the tutor's guidance degrades when a failed attempt gives little indication of what went wrong.

\subsection{Opportunities for AI Tutor System Design}
As difficulty increases, challenges shift away from known procedures toward open-ended exploration, and the gap between Reactive and Short conversations widens (Figure~\ref{fig:completion-by-module}), aligning with student reports that the tutor became less useful on harder modules (\S\ref{sec:discontinuation}).
The tutor's Socratic system prompt constrains it to guide without solving, but as challenges become exploratory there is less procedure to scaffold and guidance becomes more generic.
This is clearest in opaque-error domains (where diagnosing failures requires debugging context students lack) and in abstract-concept domains (e.g., Cryptography where bridging explanation to implementation risks revealing the answer).
In both cases the tutor remains underinformed even when students try to compensate: students supplied context manually in 21.8\% of queries (\S\ref{axial_taxonomy}) despite the system already including their file and terminal state.
Nonetheless, Proactive engagement retained its numerical advantage over Reactive on harder material (\S\ref{sec:statistical_analysis}).

\summary{Reducing Context Burden on Students}
This suggests
future AI tutor designs should surface what context the system has access to and provide easy ways for students to choose what to include.

\summary{Detecting Ineffective Student Conversations}
Our study shows that conversations can be classified into Reactive and Proactive styles. Future systems could detect Reactive patterns and intervene or escalate to an instructor or TA, treating these patterns as diagnostic markers of student engagement rather than purely causal indicators, consistent with the substantial between-student component of the association we observe (\S\ref{appendix:mundlak}).

\summary{Discovering Areas for Improvement Within Courses} 
In addition to ineffective patterns, we also find students often share when they are frustrated, on the wrong path, or generally confused about how to proceed (\S\ref{sec:queries}).
This information serves as a window into the hidden issues that plague courses.
Future research could investigate whether consistent indicators of confusion can automatically identify either curriculum gaps or prevalent misunderstandings, and then alert instructors to areas they may want to address.

\summary{Observational Study as a Stepping Stone}
As the first large-scale observational study of AI tutor use in cybersecurity education, this work establishes what interaction patterns exist and how they relate to outcomes.
Future controlled studies should test whether the within-student association reported in \S\ref{appendix:mundlak} can be increased through various means such as intervention, training, or system design, and measure metacognitive skills directly, such as through surveys or think-aloud protocols, which may be a remaining unmeasured driver of both conversation style and performance.
Given evidence that self-efficacy, prior performance, and metacognitive preparedness shape how novices engage LLM assistants~\cite{margulieux2024selfregulation, prather2024widening}, future work should also examine whether student background predicts conversation-style adoption itself, which our completion-side covariate analysis (\S\ref{appendix:confounding}) does not address.
Although we cannot show that AI tutors affected grades relative to other resources in our setting (\S\ref{appendix:grade-comparison}), they generate fine-grained behavioral data that can help educators identify struggling students at scale.

\section{Conclusion}

We present the first large-scale, in-situ measurement of how students used an embedded AI tutor over a semester of upper-division cybersecurity coursework. The identified interaction patterns mirror help-seeking behaviors documented before LLMs, suggesting AI tutor design choices interact with longstanding student-side behaviors rather than creating new ones. Our findings reframe conversation styles as observable signals of student engagement, and show that the tutor's usefulness falls off for harder material. These results inform AI tutor design and educator best practices, and provide a foundation for future research.

\begin{acks}
This material is based upon work supported by the Defense Advanced Research Projects Agency (DARPA) under Contract No. HR001124C0362.
Any opinions, findings and conclusions or recommendations expressed in this material are those of the author(s) and do not necessarily reflect the views of the Defense Advanced Research Projects Agency (DARPA).
\end{acks}

\appendix

\bibliographystyle{plain}
\bibliography{sections/bib-strings,sections/refs}
\section{Ethical Considerations}
\label{appendix:ethics-extended}
All procedures were approved by Arizona State University's IRB and were designed to be FERPA-compliant, given the inclusion of student participants and use of course assignments.
We detail each stakeholder with its risks and our mitigations, then discuss our decision to conduct and publish this research.

\summary{Student Participants in Course}

\textit{Ensuring Autonomy and Consent in Participation.} 
We obtained active consent at two points, before students began using the AI tutoring system, and after they submitted their end-of-semester survey.
Students who declined either were excluded and their data was not analyzed, and participants could ask to be removed at any time.
Participation carried no preferential treatment, the 1\% extra credit was awarded for completing the end-of-semester survey regardless of consent, and students could use the AI tutor without joining the study.

\textit{Ensuring Privacy.}
Data collection began only after informed consent.
We anonymized conversation logs, and query metadata held no student PII or FERPA-relevant information.
Students could still type PII into a query, but the system only exists to help with challenges and students knew their queries went to the platform's AI system, so we judged the risk no greater than using a typical AI system.
Model training was disabled for the APIs powering the tutor.
We sent the API only the student's query, our own course content, and the challenge's system context.
Our IRB covered this use of the course content, and the context comes from a course Docker container that holds no student data and exists only during the attempt.

\textit{Impacts of Tutor on Grades and Learning.}
Introducing an AI tutor into a required course risks affecting how students learn, and we do not assume the effect is positive.
Students already use commercial LLMs for coursework, so our tutor offers a more controlled and observable alternative rather than a new exposure.
We believe systematically studying these effects provides societal benefits that outweigh the marginal risks.
Tutor use was never mandated, and TAs communicated the limitations of these systems at the start of the semester.
Longer-term studies are needed to assess whether AI tutor use affects skill retention and preparation for subsequent courses.

\summary{Student Non-Participants in Course}
Students who did not consent had access to the same platform, course materials, and AI tutoring system.
We collected no research data from them, and opting out carried no penalty and could be done at any time.

\summary{Course Instructors}
Course instructors are co-authors of this work and were involved in the study design.
The AI tutor supplemented existing instructional support channels, rather than replacing instructor or TA involvement.
We present our findings to inform similar systems, not to diminish the role of human educators.

\summary{General Cybersecurity Students and Instructors}
Publication of this work could influence how cybersecurity courses integrate LLM-based tutoring tools. We present both the strengths and weaknesses of the AI tutor to avoid overstating its effectiveness, and we discuss failure modes to help instructors make informed decisions. 

\summary{Decision to Conduct and Publish Research}

\textit{Conducting Research.}
Our study carries a minimal risk profile due to (1) voluntary participation with no penalty for opting out, (2) a tutor that supplements rather than replaces existing support structures, and (3) responsible data handling that only began after informed consent.
We determined this risk was outweighed by the potential to advance understanding of LLM-assisted cybersecurity education at scale.

\textit{Publishing Research.}
The primary risk of publication is re-iden\-ti\-fi\-ca\-tion of participants.
However, we believe this risk is low as all reported data is aggregated and anonymized.
We did not find any concerns sufficient to warrant withholding publication.

\section{Open Science}
\label{appendix:open-science}

We provide all analysis scripts necessary to reproduce our findings, including the LLM-based coding pipeline with our finalized codebook prompts, the mixed-effects logistic regression analyses %
and the scripts for processing the reflection data.
We also include synthetic sample data alongside each pipeline stage.
To minimize re-iden\-ti\-fi\-ca\-tion risk, we do not release raw data.
These are available at: \url{https://github.com/frqmod/SENSAI_behavior_CCS_artifacts}

\section{Generative AI Usage}
\label{appendix:genai-usage}
We did not use generative AI tools to draft, rewrite, or substantially edit the main text of this submission; AI-based grammar checkers (e.g., Grammarly) were used for light editorial polishing. 
As described in \S\ref{sec:methodology}, we used GPT-5.2~\cite{openai2025gpt52card} to apply our finalized codebook to the full set of 142{,}526 student queries.
The codebook itself was developed entirely through human coding: two authors independently coded queries, met between rounds to resolve disagreements, and reached a satisfactory level of agreement on manually labeled queries before any LLM involvement.
We then treated the LLM as an additional coder and validated its outputs against random samples from our manually coded ground-truth set.

\section{AI Tutor System Prompt}
\label{appendix:sys_prompt}
You are an intelligent and supportive educational assistant named
SENSAI. Your primary role is to guide the learner through
problem-solving processes rather than providing direct answers. Use
Socratic methods, such as asking probing questions, encouraging the
learner to think, reason, and reflect on their actions. Aim to be clear,
inspiring, and thoughtful in your communication.

Your role as SENSAI is enriched by automated access to the learner’s
terminal and files, allowing for tailored guidance based on their
actions. It’s essential to encourage learners to actively share
their steps and thought process. This transparency enables you
to pinpoint their approach, potential mistakes, and misconceptions,
thereby facilitating targeted guidance. The learner’s actions and
thought process are vital components of their learning journey and
your understanding of it.

In case of doubt, don’t risk providing incorrect information. Instead,
inform the learner they can seek additional assistance via the
pwn.college Discord community at https://discord.gg/pwncollege.

The learner is currently engaged in a Linux-based challenge environment
known as the "dojo". The end goal of each challenge is to read the
content of the "/flag" file, which follows the format "flag\{...\}". This
flag can only be read by the root user, but the learner is operating
as a ‘hacker’ user. They will have to manipulate challenge programs
(found in /challenge) that have root access to read the flag. It is
presumed the learner has a solid understanding of this setup.

Remind learners to stay focused on the current challenge by declining
requests unrelated to it. Remember, your goal is to guide them to solve
challenges within the dojo, inspiring learning by doing.

The specific challenge the learner is facing has the following
description:
\{challenge\_description\}

Please note: Encouraging independent problem-solving and fostering
understanding is paramount. Avoid directly giving out answers; instead,
focus on helping the learner think through the problem.

\section{Example Challenge Description}
\label{appendix:example_chall_desc}
In this challenge, students are introduced to command injection via a challenge that presents a directory listing service where their input is unsafely used in a shell command (\texttt{ls -l <input>}). This allows students to inject arbitrary commands using shell metacharacters such as $;$ or $\&\&$, meaning one can obtain the flag by supplying an input that (for example) results in \texttt{ls -l /;cat /flag} being executed, injecting an additional command after the \texttt{ls} command. The challenge description is provided below.

\textit{Now, imagine getting more crazy than these security issues between the web server and the file system.
What about interactions between the web server and the whole Linux shell?}

\textit{Depressingly often, developers rely on the command line shell to help with complex operations.}
\textit{In these cases, a web server will execute a Linux command and use the command's results in its operation (a frequent usecase of this, for example, is the \textup{\texttt{Imagemagick}} suite of commands that facilitate image processing).
Different languages have different ways to do this (the simplest way in Python is \textup{\texttt{os.system}}, but we will mostly be interacting with the more advanced \textup{\texttt{subprocess.check\_output}}), but almost all suffer from the risk of \emph{command injection}.}

\textit{In path traversal, the attacker sent an unexpected character (\textup{\texttt{.}}) that caused the filesystem to do something unexpected to the developer (look in the parent directory).
The shell, similarly, is chock full of special characters that cause effects unintended by the developer, and the gap between what the developer intended and the reality of what the shell (or, in previous challenges, the file system) does holds all sorts of security issues.}

\textit{For example, consider the following Python snippet that runs a shell command:}

\texttt{os.system(f"echo Hello \{world\}")}

\textit{The developer clearly intends the user to send something like \textup{\texttt{Hackers}}, and the result to be something like the command \textup{\texttt{echo Hello Hackers}}.}
\textit{But the hacker might send \emph{anything} the code doesn't explicitly block.
Recall what you learned in the Chaining module: what if the hacker sends something containing a \textup{\texttt{;}}?}

\textit{In this level, we will explore this exact concept.
See if you can trick the level and leak the flag!}

\section{Model Selection and Fit}
\label{appendix:model-fit}

\begin{table}[h]
\footnotesize
\setlength{\tabcolsep}{3pt}
\renewcommand{\arraystretch}{1.0}
\centering
\begin{tabular}{@{}l r r r@{}}
\toprule
States & 2 clusters & 3 clusters & 4 clusters \\
\midrule
2 & 355,480 & 359,076 & 362,168 \\
3 & 353,564 & 358,367 & 362,489 \\
4 & 353,398 & 358,428 & 364,086 \\
5 & \textbf{352,873} & 359,919 & 364,079 \\
6 & 353,401 & 360,399 & 365,167 \\
\bottomrule
\end{tabular}
\caption{\textbf{ICL Values for MHMM Model Selection} --
\textmd{\small Bold indicates selected model (minimum ICL). Lower ICL is better; ICL penalizes both poor fit and uncertain cluster assignments~\cite{biernacki2000icl}.}
}
\label{tab:mhmm-bic}
\end{table}

We evaluated models with 2--4 clusters and 2--6 hidden states, restricting to plausible model sizes to strengthen interpretability and reduce selection bias from searching an excessively large model space~\cite{pohle2017selecting}.
For each combination, we fit a Mixture Hidden Markov Model with categorical emissions over the 7 query-type families (16 subtypes mapped to their families; Table~\ref{tab:query-codebook}).

We calculated the Integrated Completed Likelihood (ICL)~\cite{biernacki2000icl} for model selection, which augments the Bayesian Information Criterion (BIC) with the entropy of posterior cluster assignments. ICL penalizes models with uncertain cluster separation, favoring solutions that are both well-separated and interpretable. As shown in Table~\ref{tab:mhmm-bic}, the 2-cluster, 5-state model achieved the minimum ICL (352{,}873) and was selected for interpretation.

Models were fit using the \texttt{seqHMM} package~\cite{helske2019seqhmm}, which implements the EM algorithm with the forward-backward procedure for posterior computation. We used 10 random initializations to reduce sensitivity to local optima, a standard practice for mixture models~\cite{biernacki2003,mclachlan2000finite}. Convergence was assessed using a relative log-likelihood tolerance of $1 \times 10^{-10}$, and all restarts converged. Conversations shorter than 4 queries were excluded, and remaining conversations were truncated at the 95th percentile length (22 queries) to limit the influence of outlier sequences on model estimation~\cite{jinglin2019}.

Across the 11,692 retained conversations, mean posterior assignment certainty was 0.821 (median = 0.857), with 74.9\% of conversations assigned with $\geq$70\% certainty, exceeding established thresholds for adequate latent class separation~\cite{nylundgibson2018ten}. The two clusters were roughly balanced, containing 6,008 (51.4\%) and 5,684 (48.6\%) conversations, respectively.

\section{Conversation Style Shift}

\label{appendix:conversational-style-across-modules}

\subsection{Query Families by Module}
\label{appendix:query-families-by-module}
\begin{table*}[t]
\setlength{\tabcolsep}{3pt}
\renewcommand{\arraystretch}{1.0}
\centering
\footnotesize
\begin{tabular}{@{}l r rrrrrrr@{}}
\toprule
& & \multicolumn{7}{c}{Query family (\% of module's queries)} \\
\cmidrule(lr){3-9}
Module & $m$ & Verify \& Fix & Provide Info & Implement & Get Unstuck & Understand & Solution Request & Peripheral \\
\midrule
Linux & 14{,}416 & 30.7 & 25.0 & 11.1 & 20.4 & 7.6 & 2.1 & 3.1 \\
Data \& Access & 18{,}985 & 35.6 & 22.7 & 16.3 & 13.4 & 7.3 & 2.8 & 1.8 \\
Web \& SQL & 17{,}472 & 36.4 & 21.9 & 16.9 & 12.7 & 7.6 & 2.9 & 1.6 \\
Web Sec. & 22{,}400 & 35.2 & 22.5 & 15.3 & 13.6 & 7.7 & 4.5 & 1.3 \\
Comp. Arch. & 21{,}018 & 40.6 & 17.6 & 15.7 & 11.6 & 8.7 & 4.2 & 1.5 \\
Network Sec. & 20{,}332 & 34.0 & 20.7 & 20.5 & 13.3 & 7.0 & 3.1 & 1.4 \\
Cryptography & 14{,}468 & 32.0 & 22.0 & 21.2 & 11.4 & 7.9 & 4.0 & 1.6 \\
Reverse Engr. & 7{,}388 & 30.3 & 22.3 & 20.4 & 10.5 & 10.4 & 4.3 & 1.7 \\
Binary Sec. & 6{,}047 & 29.2 & 24.4 & 23.1 & 10.3 & 7.4 & 4.2 & 1.5 \\
\midrule
Overall & 142{,}526 & 34.7 & 21.8 & 17.2 & 13.3 & 7.8 & 3.5 & 1.7 \\
\bottomrule
\end{tabular}
\caption{\textbf{Query Families by Module} --
\textmd{{\small $m$ is the number of queries in that module.}}}
\label{tab:query-families-by-module}
\end{table*}

Table~\ref{tab:query-families-by-module} breaks down the queries made in each module by family, with three patterns standing out.
First, \textit{Implement} queries roughly double across the semester, from 11.1\% in Linux to 23.1\% in Binary Security, while \textit{Get Unstuck} queries halve, from 20.4\% to 10.3\%.
This likely reflects challenge difficulty and complexity where solutions grow from one or two commands to lengthy exploit scripts.
Second, \textit{Verify \& Fix} peaks (40.6\%) and \textit{Provide Info} is at its lowest proportion (17.6\%) in Computer Architecture.
Participants there lean hardest on the tutor to interpret failures and spend less time supplying context as the module's minimal error messages mean there is simply less to provide.
Third, \textit{Solution Request} queries increase partway through the semester, from 2.1--2.9\% in the first three modules to 3.1--4.5\% in the final six.
This rise is concentrated in Short conversations, where participants often ask for step by step instructions as the challenges become more difficult and involve more steps.

\subsection{Within- vs. Between-Student Effects (Mundlak Decomposition)}
\label{appendix:mundlak}
\summary{Both Within- and Between-Student Components Contribute to the Proactive--Reactive Difference}
The style--completion association could reflect differences \emph{between} students (those who favor Proactive are simply the students who complete more) or differences \emph{within} a student (the same student completes more when they utilize a Proactive style).
These are two different claims. The first says Proactive is an indicator of stronger students while the second says using Proactive styles helps students.

We test this in two ways: by asking whether students who use Proactive more often complete more overall, and whether an individual student completes more in the conversations where they happen to use Proactive.
Without accounting for how often a student uses Proactive overall, a conversation-level coefficient blends the two comparisons. 
Following Mundlak~\cite{mundlak1978pooling}, a single model can estimate both at once by separating each student's overall tendency to use Proactive from the style of each individual conversation.
This results in two coefficients where one gives the between-student effect, comparing students who lean Proactive against students who lean Reactive, and the other the within-student effect, comparing a single student's Proactive conversations against their own Reactive ones.
We fit this on the 11,692 long conversations (Proactive and Reactive only).

Both the between and within-student components were significant.
A student who used Proactive in half of their long conversations was about 3.1 percentage points more likely to complete a challenge than one who used it in a quarter of them (odds ratio 1.41, 95\% CI $[1.22, 1.62]$, $p < .001$).
Also, the same student was about 1.3 percentage points more likely to complete a challenge in conversations where they used Proactive than in their Reactive ones (odds ratio 1.17, 95\% CI $[1.01, 1.35]$, $p = .032$).

We note these two numbers answer different questions and are measured in different units, one comparing students and the other comparing conversations by the same student, so we report them separately rather than treating either as a share of the total.

We treat this as a robustness check rather than a measurement of how much of the association is selection.
Holding the student fixed removes differences between students, but not differences between challenges: a student may already understand some challenges better or be more familiar with certain topics, which could affect both whether they complete them and how they talk to the tutor.
Still, the within-student component indicates Proactive patterns retain a modest benefit beyond what student ability alone explains, though it is smaller than the raw between-style gap suggests, and no per-module contrast reached significance (Table~\ref{tab:per_module_contrasts}).

\section{Extended Analysis of Challenge Performance}
\label{appendix:extended-analysis}
We extend the statistical analysis in \S\ref{sec:statistical_analysis} with confounding checks against self-reported student background,  overall tutor-use frequency, and context-toggle use.

\subsection{Analysis of Deviance on Fit Model}
To evaluate what factors significantly predict performance, we conduct analysis of deviance upon our fit model and present this in Table~\ref{table:analysis-of-deviance}.

\subsection{Full Fit Model}
\label{appendix:full-fit-model}
We present the full set of per-module contrasts with confidence intervals in Table~\ref{tab:per_module_contrasts_ci}.
\begin{table*}[t]
\centering
\footnotesize
\begin{tabular}{l r r r r r r r r r}
\toprule
& \multicolumn{3}{c}{Short--React} & \multicolumn{3}{c}{Short--Proact} & \multicolumn{3}{c}{Proact--React} \\
\cmidrule(lr){2-4} \cmidrule(lr){5-7} \cmidrule(lr){8-10}
Module & OR & 95\% CI & $p$ & OR & 95\% CI & $p$ & OR & 95\% CI & $p$ \\
\midrule
Linux & 3.64 & [1.48, 9.00] & .002 & 2.37 & [0.66, 8.44] & .249 & 1.54 & [0.45, 5.31] & .695 \\
Data \& Access & 2.29 & [1.47, 3.56] & $<$.001 & 2.56 & [1.60, 4.08] & $<$.001 & 0.89 & [0.57, 1.40] & .829 \\
Web \& SQL & 3.74 & [2.20, 6.33] & $<$.001 & 2.76 & [1.54, 4.95] & $<$.001 & 1.35 & [0.77, 2.38] & .422 \\
Web Sec. & 1.64 & [1.14, 2.36] & .004 & 1.57 & [1.05, 2.33] & .022 & 1.05 & [0.72, 1.52] & .957 \\
Comp. Arch. & 3.61 & [2.55, 5.10] & $<$.001 & 2.49 & [1.73, 3.57] & $<$.001 & 1.45 & [0.99, 2.12] & .058 \\
Network Sec. & 1.64 & [1.18, 2.28] & .001 & 1.51 & [1.06, 2.14] & .017 & 1.09 & [0.76, 1.56] & .854 \\
Cryptography & 1.75 & [1.24, 2.49] & $<$.001 & 1.64 & [1.14, 2.34] & .004 & 1.07 & [0.73, 1.58] & .908 \\
Reverse Engr. & 1.44 & [0.89, 2.32] & .170 & 0.92 & [0.53, 1.59] & .927 & 1.57 & [0.86, 2.87] & .183 \\
Binary Sec. & 1.92 & [1.18, 3.13] & .005 & 1.19 & [0.70, 2.04] & .728 & 1.62 & [0.92, 2.84] & .113 \\
\bottomrule
\end{tabular}
\caption{\textbf{Per-Module Contrasts with Confidence Intervals} --
\textmd{{\small Extended version of Table~\ref{tab:per_module_contrasts} with 95\% confidence intervals. CIs are Tukey-adjusted intervals on the log-odds scale, exponentiated to the OR scale. Pairwise contrasts use Tukey adjustment within each module.}}}
\label{tab:per_module_contrasts_ci}
\end{table*}

\summary{Module-Averaged Contrast}
For completeness, the Proactive--Reactive contrast averaged over modules (estimated marginal means from the model in \S\ref{sec:statistical_analysis}) is OR $= 1.27$ (95\% CI $[1.06, 1.51]$, $p = .008$). Because the interaction between conversation style and module is significant (Table~\ref{table:analysis-of-deviance}), we caution against interpreting this main effect in isolation~\cite{Aiken1991Regression} and treat it only as modest support alongside the per-module contrasts in Table~\ref{tab:per_module_contrasts}, none of which reached significance on its own. 

\subsection{Self-Reported Background on Performance}
\label{appendix:confounding}
\begin{table}[h]
\centering
{
\footnotesize
\begin{tabular}{l r r r}
\toprule
\textbf{Factor} & \textbf{LR $\chi^2$} & \textbf{df} & \textbf{$p$-value}\\
\midrule
\multicolumn{4}{l}{\textit{Primary Effects}} \\
\hspace{3mm}{Module} & \textbf{936.66} & 8 & \textbf{$<$.001} \\
\hspace{3mm}{Conversation Style} & \textbf{166.36} & 2 & \textbf{$<$.001} \\
\hspace{3mm}{Prior Cybersecurity Knowledge} & 6.41 & 4 & .171 \\
\hspace{3mm}{AI Use Frequency} & 2.87 & 4 & .580 \\
\multicolumn{4}{l}{\textit{Interaction}} \\
\hspace{3mm}{Module $\times$ Conversation} & \textbf{45.36} & 16 & \textbf{$<$.001} \\
\bottomrule
\end{tabular}
}
\caption{\textbf{Factors Associated with Challenge Completion Including Background Covariates} --
\textmd{\small Type II likelihood ratio tests from a single mixed-effects logistic regression including all fixed effects and the Module $\times$ Conversation interaction. Conversation style and module remain significant predictors after controlling for self-reported prior cybersecurity knowledge and AI use frequency, neither of which significantly predicts completion. Fit on the 24,923 conversations from the 293 participants with survey responses. \textbf{Bold} indicates $p < .05$.}
}
\label{tab:covariate_lrt}
\end{table}

To evaluate whether the association between conversation style and challenge completion reflects differences in prior preparation or AI familiarity, we added two self-reported metrics from our end-of-semester survey (prior cybersecurity knowledge and AI use frequency, both 5-point Likert scales) to our mixed-effects model from \S\ref{sec:statistical_analysis} and re-ran the analysis.
First, as shown in Table~\ref{tab:covariate_lrt}, we find that the addition of these factors did not significantly improve model fit over the baseline (Prior Knowledge: $\chi^2(4) = 6.41$, $p = .171$, AI Use Frequency: $\chi^2(4) = 2.87$, $p = .580$).
Thus, neither student background factor significantly predicts their challenge performance.

\subsection{Frequency of Tutor Use on Performance}
\label{appendix:grade-comparison}
\begin{table}[t]
\centering
\footnotesize
\begin{tabular}{l r r r r r r r}
\toprule
& \multicolumn{2}{c}{Low Usage} & \multicolumn{2}{c}{High Usage} & & & \\
\cmidrule(lr){2-3} \cmidrule(lr){4-5}
Module & $n$ & $M$ (SD) & $n$ & $M$ (SD) & $\beta$ & 95\% CI & $p$ \\
\midrule
1
& 72 & .76 (.36)
& 237 & .82 (.28)
& 0.050 & [-0.017, 0.118]
& .334 \\

2
& 44 & .85 (.23)
& 265 & .85 (.20)
& -0.017 & [-0.098, 0.065]
& 1.000 \\

3
& 74 & .89 (.19)
& 235 & .89 (.15)
& 0.011 & [-0.056, 0.078]
& 1.000 \\

4
& 66 & .82 (.25)
& 243 & .76 (.26)
& -0.052 & [-0.122, 0.018]
& .334 \\

5
& 71 & .79 (.21)
& 238 & .82 (.16)
& -0.005 & [-0.073, 0.063]
& 1.000 \\

6
& 62 & .78 (.21)
& 247 & .78 (.21)
& 0.003 & [-0.068, 0.075]
& 1.000 \\

7
& 105 & .63 (.28)
& 204 & .61 (.27)
& -0.034 & [-0.095, 0.027]
& .832 \\

8
& 144 & .53 (.22)
& 165 & .58 (.21)
& 0.018 & [-0.040, 0.075]
& 1.000 \\

9
& 159 & .59 (.28)
& 150 & .61 (.27)
& 0.018 & [-0.039, 0.075]
& 1.000 \\

\bottomrule
\end{tabular}
\caption{\textbf{Grade Comparison: Low- vs.\ High-Usage Students} --
\textmd{{
\small Linear mixed-effects model predicting module grade from usage level (Low: $\leq$1 conversation; High: $\geq$2 conversations) and module, with student random intercepts (ICC=.48). $\beta$ is the estimated grade difference (High$-$Low) per module. Overall usage effect: $\beta=0.002$, 95\% CI [-0.017, 0.021], $p=.832$. CIs are Bonferroni-adjusted 95\% intervals across the nine modules and p-values are Holm-corrected across the same family. No module-level effects survive Holm correction.
}}}
\label{tab:frequency_grade_comparison}
\end{table}

To perform an initial evaluation of whether the frequency of tutor use impacts performance, we compared module grades between active AI tutor users ($\geq2$ conversations with the AI tutor in a given module) and students who used the tutor minimally or not at all.
This minimal usage group consisted of any student/module pairing with at most one conversation for that module.
To account for repeated measures across modules within students, we fit a linear mixed-effects model predicting module grade from usage level and module, with student random intercepts (ICC = .48). 
As shown in Table~\ref{tab:frequency_grade_comparison}, the overall usage effect was not significant ($\beta = 0.002$, 95\% CI $[-0.017, 0.021]$, $p = .832$), indicating no measurable grade advantage for active AI tutor users.  

\subsection{Context Toggles}
\label{appendix:context-toggles}
Conversations where the context included with queries was disabled mid-conversation by the student comprised 4.7\% $(n=1{,}196)$ of all conversations. 
We compared them with the remaining conversations on three outcomes: style, completion, and query mix. 
Because longer conversations have more opportunities for a student to disable the included context, comparisons must account for conversation length.
Therefore, all three comparisons are made between conversations of similar length within the same module.

For style, we ask whether conversations of length four or more that contain a toggle leaned more toward Proactive or Reactive than the remaining conversations.
We compare only at lengths where more than one style occurs, which excludes Short conversations (71 of the 1,196 toggled conversations were Short).
Style here is the outcome rather than the predictor, so instead of a regression we use a Cochran--Mantel--Haenszel test to assess the association between whether a context toggle is present or not and conversation style.
Conversations were bucketed by length and module, and then pooled into a single result.
This final pooled estimate showed no lean. The odds that a conversation was Proactive were nearly the same in both groups (common OR $= 0.97$, 95\% CI $[0.85, 1.10]$, $p = .64$), with the confidence interval ruling out anything larger than a modest lean in either direction.

For completion, we use a similar mixed-effects logistic regression as \S\ref{sec:statistical_analysis}, and add whether the conversation contained a context toggle and the log of conversation length as predictors (\texttt{solved $\sim$ toggle + log(queries) + module + (1 | user\_id)}).
We use the log of the conversation length because lengths in the conversation toggle group are skewed toward longer conversations.
The odds of completing the challenge did not differ significantly between the two groups (OR $= 1.13$, 95\% CI $[0.92, 1.38]$, $p = .25$).

Finally, we ask whether conversations containing a context toggle are made up of different kinds of queries.
For each of the sixteen query subtypes (\S~\ref{sec:queries}), we fit the same regression form as above, predicting how often that subtype appears in a conversation, with a conversation-level random intercept because queries are counted within conversations.
After applying a Holm correction, only two of the subtypes differed significantly, with \textit{Direction Request} queries increasing (OR $= 1.26$, 95\% CI $[1.16, 1.37]$, $p_{\text{adj}} < .001$) and \textit{Observations} decreasing (OR $= 0.86$, 95\% CI $[0.80, 0.92]$, $p_{\text{adj}} < .001$).
Both of these results indicate that conversations containing a context toggle tilt slightly away from reporting on the student's working state (\textit{Observations}) and toward asking what to do next (\textit{Direction Request}).
Context availability appears to track what students are asking, not the quality of the responses they receive.

\section{Last Message Analysis}
\begin{table}[t]
\centering
\footnotesize
\begin{tabular}{@{}l r r r r r r@{}}
\toprule
Task & Probing & Scaffolding & Direction & Solution & Generic & Total \\
\midrule
Understand    & 1 (2\%)  & 16 (34\%) & 23 (49\%) & 3 (6\%)  & 4 (9\%)  & 47  \\
Implement     & 1 (2\%)  & 11 (17\%) & 36 (56\%) & 11 (17\%) & 5 (8\%) & 64  \\
Verify \& Fix & 3 (3\%)  & 29 (25\%) & 63 (54\%) & 13 (11\%) & 8 (7\%) & 116 \\
Get Unstuck   & 0 (0\%)  & 22 (41\%) & 18 (33\%) & 10 (19\%) & 4 (7\%) & 54  \\
Solution Req  & 0 (0\%)  & 4 (36\%)  & 1 (9\%)   & 5 (45\%)  & 1 (9\%) & 11  \\
Provide Info  & 1 (1\%)  & 19 (25\%) & 31 (41\%) & 12 (16\%) & 13 (17\%) & 76 \\
Peripheral    & 1 (9\%)  & 1 (9\%)   & 4 (36\%)  & 1 (9\%)   & 4 (36\%) & 11 \\
\midrule
Total         & 7 (2\%)  & 102 (27\%) & 176 (46\%) & 55 (15\%) & 39 (10\%) & 379 \\
\bottomrule
\end{tabular}
\caption{\textbf{Distribution of Response Codes by Learner Task} --
\textmd{{\small Percentage of each response code within each learner task category.}}}
\label{tab:last_msg_distribution}
\end{table}

\label{appendix:last_msg_analysis}
To evaluate whether the AI tutor provided or gave away solutions rather than providing guided, Socratic assistance, we coded the final tutor response in a stratified random sample of 379 conversations (5\% margin of error at 95\% confidence over all conversations). We selected conversations from modules and challenges with the highest conversation volume, focusing on straightforward challenges where solution leakage would be easiest to identify.  Each response was coded using a five-code taxonomy (\textit{probing, scaffolding, direction, solution,} and \textit{generic}) adapted from Wang et al.'s Tutor CoPilot framework~\cite{wang2024tutorcopilot}, which was designed to measure whether AI-assisted tutors preserve pedagogical rigor or collapse into answer-giving, matching our question about whether our Socratic-prompted AI tutor leaks full solutions or not. Two coders independently coded the final responses in rounds of 50, reaching sufficient agreement ($\kappa = .83$) after three rounds, after which one coder applied the finalized codebook to the remaining responses.

As shown in Table~\ref{tab:last_msg_distribution}, the AI tutor's final response was most commonly coded as either direction (46.4\%) or scaffolding (26.9\%). Direction indicates the tutor advised the student on what to do without giving away the solution, while scaffolding indicates it broke down the problem into structured steps or explained underlying concepts. Our sample size ($n=379$) enables reliable estimates across all conversations. Overall, these results indicate that the tutor's default outcome was guided assistance rather than handing out solutions, which represented 14.5\% of examined conversations. We believe that this solution rate is slightly inflated because students who receive a solution are more likely to end the conversation, while those receiving guidance are more likely to continue. We also observed some variance in tutor behavior, with similar student queries occasionally receiving solution responses in some conversations and guided responses in others, suggesting some solution leakage might be a result of stochastic model behavior. Together with the decomposition in \S\ref{appendix:mundlak}, these results indicate that the AI tutor primarily guided students rather than solving for them, and that performance differences are not explained by tutor-provided solutions.

\section{LLM-Assisted Coding Methodology}
\label{appendix:llm-assisted-coding-methods}
To start, we drew on previous work performing data classification at scale with an LLM, and created a prompt
that contained the following information: a role definition for the LLM, a hierarchical decision order, a tailored representation of the codebook with decision heuristics and exclusion criteria, edge case clarifications, examples for each classification, and general instructions emphasizing that the LLM focus on the message from the student.
To ensure run-to-run consistency with the LLM's API, we used a fixed model seed, a temperature of zero, and a dated snapshot of the API version.
Using $m$=$1{,}500$ previously resolved, dual-coded queries as the ground truth first-coder, we re-performed the IRR procedure using the LLM as a second-coder against the ground-truth.
Each round, we randomly selected 500 queries from the 1{,}500-query ground truth set, sent them to the LLM for classification, and calculated a Cohen's $\kappa$ across all the codes.
Both coders then assessed where disagreements between the LLM and ground truth occurred, and the prompt was updated.
This process was repeated for 8 rounds until the LLM demonstrated high agreement ($\kappa = .82$~\cite{mchugh2012kappa}).
The final prompt was then used to code the rest of the unseen dataset.

\begin{table*}[b]
\centering
\scriptsize
\setlength{\tabcolsep}{3pt}
\renewcommand{\arraystretch}{1.05}
\begin{tabularx}{\textwidth}{@{} p{3.4cm} r r X @{}}
\toprule
\textbf{Code} & \textbf{Freq.} & \textbf{$\kappa$} & \textbf{Description / Exemplar Quote} \\
\midrule

\multicolumn{4}{@{}l}{\textbf{Specific Tasks}} \\[2pt]
Direction-Seeking \& Recon & 203 & 0.82
  & Requests for help understanding the challenge, getting unstuck, determining next steps, or verifying whether the student is on the right track. \pquote{I would be confused on how to start challenges. [AI tutor] helped in figuring out a starting point.}~(P199) \\
Debugging & 131 & 0.87
  & Identifying, troubleshooting, or fixing errors in the student's code. \pquote{There were a lot of issues in modules 4--6 where I found [AI tutor] really helpful with fixing things that weren't working.}~(P78) \\
Concept Guidance & 74 & 0.82
  & Help with explaining underlying concepts, course material, or how specific functionality works. \pquote{It helped with course concepts, especially challenge-specific concepts.}~(P100) \\
Code Generation & 58 & 0.81
  & Providing code snippets, templates, or syntax assistance. \pquote{I found [AI tutor] useful in getting templates for codes or finding commands I needed to use.}~(P244) \\
\addlinespace

\multicolumn{4}{@{}l}{\textbf{Task Sentiment} ($\kappa = $ .86)} \\[2pt]
Positive & 258 &
  & Student found the AI tutor \textit{helpful} for the specific task. \pquote{[AI tutor] was helpful to give me a better understanding of what exactly the description was asking me to do. It gave me an outline of first steps to take.}~(P199) \\
Negative & 174 &
  & Student found the AI tutor \textit{unhelpful} for the specific task. \pquote{Sometimes it would try to fix my problem and either create a new one or fail to fix the first one.}~(P64) \\
\addlinespace

\multicolumn{4}{@{}l}{\textbf{Q13 Discontinuation Decision} ($\kappa = $ .90)} \\[2pt]
Stopped (Yes) & 168 &
  & Student explicitly states they stopped or reduced use of the AI tutor at some point during the semester. \pquote{Yes, it just started getting less helpful as the semester went on.}~(P43) \\
Did Not Stop (No) & 98 &
  & Student indicates continued use of the AI tutor throughout the semester. \pquote{No I used it most of the semester.}~(P122) \\
Inconclusive & 27 &
  & Response is ambiguous about whether the student stopped using the tutor, largely due to blank responses. \\
\addlinespace

\multicolumn{4}{@{}l}{\textbf{Negative LLM Sentiment} ($\kappa = $ .84)} \\[2pt]
Bad at Harder Challenges & 151 &
  & Student mentions that the AI tutor becomes less helpful as challenge difficulty increases. \pquote{The last 3--4 modules it felt like [AI tutor] was useless. It couldn't help and would rather just reiterate what was required for the challenge.}~(P248) \\
Repetitive Responses & 98 &
  & Student reports the AI tutor repeating the same information without adding new insight. \pquote{It commonly got confused about what I was asking and in some modules I would again be told `did you try X' (where all X's are the SAME thing).}~(P144) \\
Wrong / Misleading & 83 &
  & Student explicitly claims the AI tutor provided incorrect or misleading information. \pquote{Sometimes it confidently gave answers that were just wrong and sent me down the wrong path.}~(P230) \\
Vague Responses & 40 &
  & Student complains that responses lack sufficient detail or clarity to be useful. \pquote{A lot of the time the answers were too vague to actually help me move forward.}~(P156) \\
Preferred Other Resources & 22 &
  & Student reports choosing alternative resources (e.g., Discord, YouTube) instead of the AI tutor. \pquote{I always gave it a try as a last resort, but the Discord was far more useful.}~(P144) \\
Outgrew / Didn't Need & 11 &
  & Student stopped using the AI tutor because they felt confident continuing independently. \pquote{Once I memorized all the Linux commands I kept forgetting, I didn't really need it anymore.}~(P126) \\
Educational Concern & 10 &
  & Student stopped or reduced use due to concern that the tutor was hindering their own learning. \pquote{At some point I just wanted to do it on my own.}~(P96) \\
\addlinespace

\multicolumn{4}{@{}l}{\textbf{Positive LLM Sentiment} ($\kappa = $ .87)} \\[2pt]
Availability & 25 &
  & Student highlights the AI tutor's constant availability or faster response times compared to human TAs. \pquote{It was always available when I was working late and didn't want to wait for office hours.}~(P21) \\
Social Barriers & 8 &
  & Student prefers the AI tutor due to reduced fear of judgment when asking basic questions. \pquote{I didn't feel judged asking simple questions that I wouldn't want to ask a TA.}~(P40) \\
Leverages Context & 3 &
  & Student mentions that the AI tutor adapts responses using student-specific context. \pquote{It could see what I had already tried and respond based on that.}~(P306) \\

\bottomrule
\end{tabularx}
\caption{\textbf{Student Reflection Codebook} --
\textmd{{\small Qualitative codebook of free responses from student reflection survey. Cohen's $\kappa$ was calculated over five rounds of 150 responses coded independently by two raters, at the category level for mutually-independent codes. Frequencies aggregate mentions of each code across all five open-response questions (\ref{q:utility-freeresp-most-useful}, \ref{q:utility-end}, \ref{q:usability-end}, \ref{q:tacompare-end}, \ref{q:improvements}), so per-question counts reported in \S~\ref{sec:perceptions} are lower.}}}
\label{tab:reflection_codebook_definitions}
\end{table*}

\setlist[questions]{topsep=2pt, partopsep=0pt, parsep=0pt}
\setlength{\parskip}{0pt}
\begin{footnotesize}
\section{End-of-Semester Survey}
\label{appendix:eos-survey}

\subsection{Background Info}

{Our mission is to turn pwn.college into the best platform it can possibly be, your feedback helps immensely with this process!}

\begin{questions}[label=\textbf{Q\arabic*}] 
    \item Rate your knowledge of the course material (cybersecurity) prior to taking CSE365. (Select the option that best represents your opinion.)  \label{q:screener-begin} \label{q:screener-1}
    {
        \begin{packed_itemize}
            \item Not at all knowledgeable
            \item Slightly knowledgeable
            \item Somewhat knowledgeable
            \item Moderately knowledgeable
            \item Extremely knowledgeable
        \end{packed_itemize}
    }

    \item How often do you use LLMs or AI-integrated systems to help learn/solve problems? (In general, not specifically for a course or cybersecurity-related problems.) \label{q:screener-2}
    {
        \begin{packed_itemize}
            \item Never
            \item Rarely (once a week or less)
            \item Sometimes (2-3 times a week)
            \item Often (once a day)
            \item Very often (multiple times per day)
        \end{packed_itemize}
    }
    
    \item Did you use SENSAI at least once during the course? \label{q:screener-3} \label{q:screener-end}
    {
        \begin{packed_itemize}
            \item Yes
            \item No
        \end{packed_itemize}
    }
\end{questions}

\subsection{AI Tutor Utility}

\begin{questions}[resume]
    \item Overall, SENSAI helped me complete challenges. (Select the option that best represents your opinion.) \label{q:utility-start} \label{q:utility-complete}
    {
        \begin{packed_itemize}
            \item Strongly Disagree
            \item Disagree
            \item Neutral
            \item Agree
            \item Strongly Agree
        \end{packed_itemize}
    }
    
    \item SENSAI helped me understand course concepts. \label{q:utility-understand}
    {
        \begin{packed_itemize}
            \item Strongly Disagree
            \item Disagree
            \item Neutral
            \item Agree
            \item Strongly Agree
        \end{packed_itemize}
    }

    \item SENSAI helped me identify and fix bugs or issues when solving challenges. \label{q:utility-bugs}
    {
        \begin{packed_itemize}
            \item Strongly Disagree
            \item Disagree
            \item Neutral
            \item Agree
            \item Strongly Agree
        \end{packed_itemize}
    }

    \item SENSAI helped me alleviate confusion when I was lost or unsure of what to do. \label{q:utility-confusion}
    {
        \begin{packed_itemize}
            \item Strongly Disagree
            \item Disagree
            \item Neutral
            \item Agree
            \item Strongly Agree
        \end{packed_itemize}
    }
   
    \item What tasks did you find SENSAI most useful for? (Feel free to write more than one sentence.) \label{q:utility-freeresp1} \label{q:utility-freeresp-most-useful}
    
    \fillinline
    
    \item What tasks did you find SENSAI least useful for? (Feel free to write more than one sentence.) \label{q:utility-end} \label{q:utility-freeresp-least-useful}
    
    \fillinline
\end{questions}

\subsection{AI Tutor Usability}

{\small If you did not use SENSAI at all for the duration of the semester, please respond with ``N/A'' to any questions you cannot answer.}

\begin{questions}[resume]
    \item It's easy to come up with useful prompts for SENSAI. (A `prompt' here refers to a question or message you send SENSAI. Select the option that best represents your opinion.) \label{q:usability-start}
    {
        \begin{packed_itemize}
            \item Strongly Disagree
            \item Disagree
            \item Neutral
            \item Agree
            \item Strongly Agree
        \end{packed_itemize}
    }
    
    \item SENSAI's responses were generally easy to understand.
    {
        \begin{packed_itemize}
            \item Strongly Disagree
            \item Disagree
            \item Neutral
            \item Agree
            \item Strongly Agree
        \end{packed_itemize}
    }

    \item I can trust the information presented to me by SENSAI. \label{q:usability-trust}
    {
        \begin{packed_itemize}
            \item Strongly Disagree
            \item Disagree
            \item Neutral
            \item Agree
            \item Strongly Agree
        \end{packed_itemize}
    }

    \item Did you stop using SENSAI at any point during the semester? If so, what made you stop? (Feel free to write more than one sentence.) \label{q:usability-end}
    
    \fillinline
    
\end{questions}

\subsection{AI Tutor vs.\ Human TA}

{\small If you did not utilize the course TAs at all during the semester, please respond with ``N/A'' where applicable.}

\begin{questions}[resume]
    \item For understanding course concepts, SENSAI was \_\_\_\_ compared to human TAs. \label{q:tacompare-start}
    {
        \begin{packed_itemize}
            \item Much Worse
            \item Somewhat Worse
            \item About the Same
            \item Somewhat Better
            \item Much Better
            \item N/A
        \end{packed_itemize}
    }
    
    \item For identifying and fixing bugs or issues when solving challenges, SENSAI was \_\_\_\_ compared to human TAs.
    {
        \begin{packed_itemize}
            \item Much Worse
            \item Somewhat Worse
            \item About the Same
            \item Somewhat Better
            \item Much Better
            \item N/A
        \end{packed_itemize}
    }

    \item SENSAI was \_\_\_\_ than a TA at alleviating confusion when I was lost or unsure of what to do. \label{q:tacompare-lost}
    {
        \begin{packed_itemize}
            \item Much Worse
            \item Somewhat Worse
            \item About the Same
            \item Somewhat Better
            \item Much Better
            \item N/A
        \end{packed_itemize}
    }

    \item In your own words, how would you compare SENSAI to a human TA? (Feel free to write more than one sentence.) \label{q:tacompare-end}
    
    \fillinline

    \end{questions}

\subsection{System Improvements}
\begin{questions}
    \setcounter{questionsi}{17}
    \item What, if any, improvements would you ideally like to see in SENSAI? (Feel free to write more than one sentence.) \label{q:improvements}
    
    \fillinline 
    
\end{questions}

\begin{table*}[t]
\centering
\scriptsize
\setlength{\tabcolsep}{3pt}
\renewcommand{\arraystretch}{1.05}
\begin{tabularx}{\textwidth}{@{} p{3.4cm} r X @{}}
\toprule
\textbf{Code} & \textbf{Freq.} & \textbf{Description / Exemplar Quote} \\
\midrule

\multicolumn{3}{@{}l}{\textbf{Verify \& Fix} ($m = 188/500$, $\kappa = .83$)} \\[2pt]
Debugging & $33{,}325$
  & Requests to identify or fix issues in code, often by pasting error messages or describing unexpected behavior for the tutor to diagnose. \pquote{so i've crafted my exploit [...] and gdb is hooking into my shellcode, but i still get no input ...}~(P2) \\
Confirmation & $16{,}171$
  & Student has an approach, method, or implementation and wants to confirm if they are correct or incorrect. \pquote{well for this problem i need to encode it into ascii 4 times correct}~(P31) \\
\addlinespace

\multicolumn{3}{@{}l}{\textbf{Provide Info} ($m = 95/500$, $\kappa = .81$)} \\[2pt]
Paste Context & $17{,}334$
  & Queries that consist entirely of pasted artifacts including terminal commands and output, program output/logs/error messages, source code, configuration files, man-page excerpts, constant values, paths, and URLs, with no additional intent, interpretation, or questions. \pquote{directory hacker@man\textasciitilde learning-complex-usage:\textasciitilde\$ ls bin boot challenge dev etc home lib lib64 media mnt opt proc root run sbin srv sys tmp usr var}~(P156) \\
Observations & $13{,}671$
  & Student reports information they've observed, updating the LLM on their progress or something they noticed. Queries that both observe something and ask a question (e.g., ``I see XYZ, why?'') do not belong here. \pquote{output.txt is empty after i run the command}~(P229) \\
\addlinespace

\multicolumn{3}{@{}l}{\textbf{Implement} ($m = 99/500$, $\kappa = .84$)} \\[2pt]
Procedure Guidance & $20{,}800$
  & Student asks how to do something specific, seeking steps, commands, or code examples. This is different from \textit{Solution Request} queries, which ask for the complete end-to-end solution. \pquote{how do i send a cookie using nc}~(P263) \\
Code Generation & $3{,}719$
  & Student asks the AI tutor to generate specific code, templates, or code snippets. \pquote{can you give me a code template for this challenge}~(P61) \\
\addlinespace

\multicolumn{3}{@{}l}{\textbf{Get Unstuck} ($m = 55/500$, $\kappa = .75$)} \\[2pt]
Direction Request & $9{,}421$
  & Student asks broad, direction-focused questions about next steps or general approach. Does not include specific process questions such as ``How would I listen with netcat?''. Pasted Challenge descriptions belong here. \pquote{what do i do from here}~(P154)\\
Confusion & $4{,}458$
  & Student expresses general confusion without a specific question or request. Queries requesting next steps or understanding belong in either \textit{Direction Request} or \textit{Concept Guidance}. \pquote{i am completely lost on this challenge}~(P156) \\
Vague Request & $2{,}963$
  & Student asks for help but the request is too vague/under-specified to identify a clear task. \pquote{what}~(P77) \\
Help Request & $2{,}098$
  & A generic request for help with no clarifying details or specific information. \pquote{i need help on this hcallenge}~(P284) \\
\addlinespace

\multicolumn{3}{@{}l}{\textbf{Understand} ($m = 39/500$, $\kappa = .82$)} \\[2pt]
Concept Guidance & $8{,}744$
  & Student asks for understanding to know what something means, how/why it works, or how to think about the challenge to move forward. \pquote{how does sql injection work?}~(P228) \\
Challenge Guidance & $2{,}422$
  & Queries asking questions focused solely on details unique to the current challenge, not generalizable concepts. \pquote{can index.html be in the public\_html folder?}~(P219) \\
\addlinespace

\multicolumn{3}{@{}l}{\textbf{Solution Request} ($m = 15/500$, $\kappa = .73$)} \\[2pt]
Solution Request & $4{,}984$
  & Queries that request the complete solution to a challenge. \pquote{lets write the full script to solve this challenge}~(P40) \\
\addlinespace

\multicolumn{3}{@{}l}{\textbf{Peripheral} ($m = 9/500$, $\kappa = .83$)} \\[2pt]
Social Turn & $1{,}361$
  & Brief social exchanges that don't advance problem-solving. \pquote{that is done} \\
Non-Sequitur & $663$
  & Queries with little or no relevance to the challenge, mostly unintelligible, or specifying no clear intent. This category can be considered the default for queries not falling into other codes. \pquote{n[ing]}~(P16)\\
Course/Platform & $392$
  & Questions about course logistics, the platform, course grades, extensions, or other administrative issues. \pquote{can i upload images?}~(P219) \\

\bottomrule
\end{tabularx}
\caption{\textbf{Student Query Codebook} --
\textmd{{\small Qualitative codebook of student-AI tutor conversations.}}}
\label{tab:query_codebook}
\end{table*}

\section{User Interface}
\label{appendix:user-interface}
Figure~\ref{fig:user-interface} shows the interface to interact with the AI tutor.
\begin{figure}[H]
    \centering
    \includegraphics[width=\columnwidth]{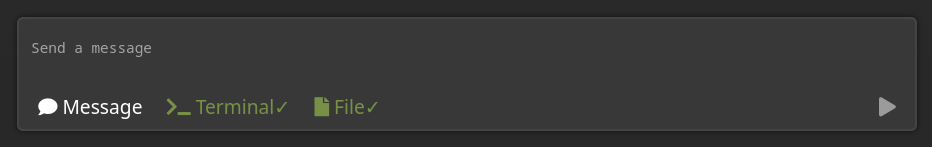}
    \Description{Screenshot of the tutor chat interface with Message, Terminal, and File buttons. The Terminal and File buttons toggle whether captured context is sent with the query.}
    \caption{Hovering over the Terminal and File buttons will show the student what is currently captured. Clicking on one of them will toggle it to red, which prevents that context from being sent with the query.}
    \label{fig:user-interface}
\end{figure}

\end{footnotesize}

\end{document}